\documentclass[preprint,12pt]{elsarticle}

\RequirePackage{natbib}
\usepackage{comment}
\usepackage{mynotes}
\usepackage{amsmath,amsthm,amsfonts}
\usepackage{bbm}
\usepackage{hyperref}
\hypersetup{
    colorlinks=true,
    linkcolor=blue,
    urlcolor=blue,
    citecolor=blue
}
\DeclareMathOperator{\erfc}{erfc}

\usepackage{amssymb}

\journal{Journal of Econometrics}

\begin{document}

\begin{frontmatter}



\title{Robust Nonparametric Stochastic Frontier Analysis}


\author[IHME,HMS]{Peng Zheng}

\affiliation[IHME]{organization={Institute for Health Metrics and Evaluation},
            addressline={University of Washington}, 
            city={Seattle},
            postcode={98105}, 
            state={WA},
            country={USA}}

\author[Amath]{Nahom Worku}

\author[Biostat]{Marlena Bannick}

\author[IHME,HMS]{Joseph Dielemann}

\author[IHME,HMS]{Marcia Weaver}

\author[IHME,HMS]{Christopher Murray}

\author[IHME,HMS,Amath]{Aleksandr Aravkin}
            
\affiliation[HMS]{organization={Department of Health Metrics Sciences},
            addressline={University of Washington}, 
            city={Seatle},
            postcode={98105}, 
            state={WA},
            country={USA}}

\affiliation[Amath]{organization={Department of Applied Mathematics},
            addressline={University of Washington}, 
            city={Seatle},
            postcode={98105}, 
            state={WA},
            country={USA}}

\affiliation[Biostat]{organization={Department of Biostatistics},
            addressline={University of Washington}, 
            city={Seatle},
            postcode={98105}, 
            state={WA},
            country={USA}}

\begin{abstract}
Benchmarking tools, including stochastic frontier analysis (SFA), data envelopment analysis (DEA), and its stochastic extension (StoNED) are core tools in economics used to  estimate an efficiency envelope and production inefficiencies from data. The problem appears in a wide range of fields -- for example, in global health the frontier can quantify efficiency of interventions and funding of health initiatives.

Despite their wide use, classic benchmarking approaches have key limitations that preclude even wider applicability. Here we propose a robust non-parametric stochastic frontier meta-analysis (SFMA) approach that fills these gaps. First, we use flexible basis splines and shape constraints to model the frontier function, so specifying a functional form of the frontier as in classic SFA is no longer necessary. Second, the user can specify relative errors on input datapoints, enabling population-level analyses. Third, we develop a likelihood-based trimming strategy to robustify the approach to outliers, which otherwise break available benchmarking methods. 

We provide a custom optimization algorithm for fast and reliable performance.  We implement the approach and algorithm in an open source Python package `sfma'. Synthetic and real  examples show the new capabilities of the method, and are used to compare SFMA to state of the art benchmarking packages that implement DEA, SFA, and StoNED.

\end{abstract}

\begin{graphicalabstract}
\includegraphics[scale=0.5]{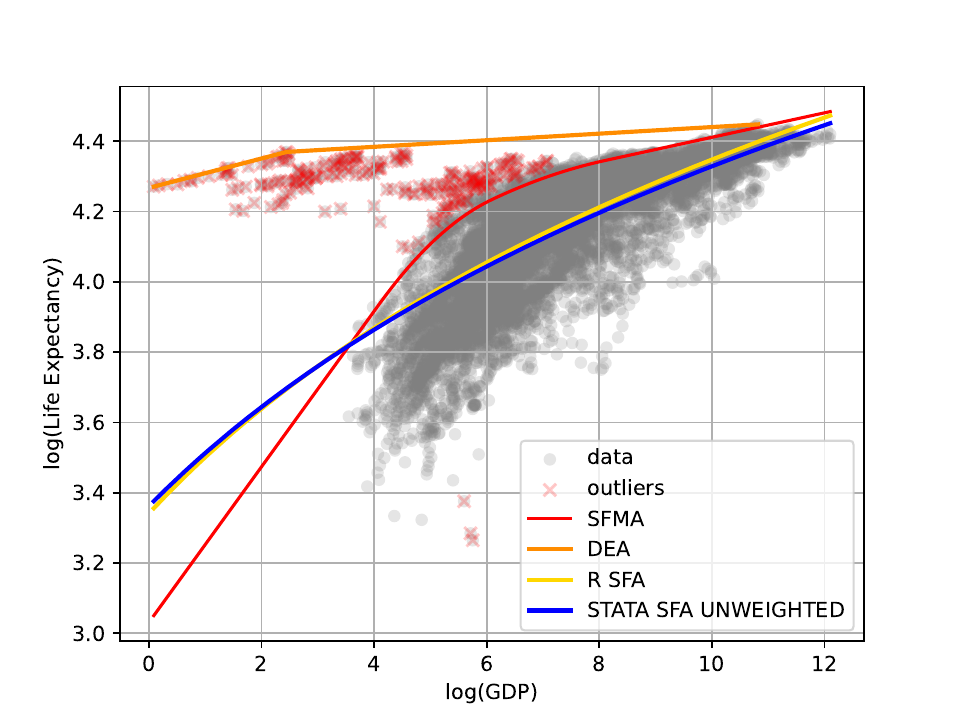}
Benchmarking tools, including stochastic frontier analysis (SFA) and data envelopment analysis (DEA) are used to  estimate an efficiency envelope and production inefficiencies from data. The problem appears in a wide range of fields -- in the selected graphic above, the the frontier can be used to quantify how efficiently countries use their GDP to extend life expectancy. 

We propose a robust non-parametric stochastic frontier meta-analysis (SFMA) approach. We use flexible basis splines and shape constraints to model the frontier function, allow the user to incorporate reported errors for input datapoints, and develop a likelihood-based trimming strategy to robustify the approach to outliers. 

We provide a custom optimization algorithm for fast and reliable performance, and distribute the resulting method in the open source Python \verb{sfma{ package. Synthetic and real examples show SFMA is competitive compared to state of the art benchmarking packages in R, Python, and Stata. 
\end{graphicalabstract}

\begin{highlights}
\item We present a new semi-parametric benchmarking approach that generalizes stochastic frontier analysis (SFA) to allow flexible spline modeling of the frontier. 

\item The meta-analytic focus allows the SFMA model to incorporate and estimate different statistical errors, including reported sampling error, as well as unknown non-sampling error and inefficiencies. 

\item SFMA also features trimming functionality to remove outliers, that helps to obtain useful fits in challenging real-world cases. 

\item The SFMA package is available for general public use through the Python package \verb{sfma{, available at \url{https://github.com/ihmeuw-msca/sfma}.
\end{highlights}



\end{frontmatter}


\section{Introduction}

A  goal in many econometric analysis is to estimate the\textit{ production function }-- the maximum amount of output obtainable from a set of inputs to a production process -- and the associated \textit{firm}-specific production inefficiencies. The framework is highly versatile, with a myriand benchmarking applications in the energy industry~\cite{filippini2014applications,lin2015stochastic}, agriculture~\cite{benedetti2019evaluating,bibi2021technical}, education~\cite{sahnoun2020education,titus2016examining}, and manufacturing~\cite{charoenrat2014efficiency,onder2003efficiency}. 
In the global health context, the output may be a measure of healthcare access, the inputs may be investments in healthcare, and inefficiencies may be country-specific gaps between the ideal production function and observations~\cite{lozano2020measuring}. 
Current approaches to the problem can be classified along the lines of envelopment through linear programming, exemplified by DEA, its stochastic extension StoNED, and likelihood-based approaches such as SFA. 

The approach in this paper, dubbed SFMA, addresses three key aspects:
\begin{itemize}
\item Nonparametric modeling of the frontier using splines
\item Detailed modeling of statistical errors, including reported sampling error, as well as unknown non-sampling error and inefficiencies 
\item Automated outlier detection and removal. 
\end{itemize}
To provide a robust software tool, we implement a customized optimization algorithm that exploits problem structure and performs a single unified analysis that determines outliers, estimates error statistics, and infers the spline coefficients of the frontier and other covariate multipliers of interest. 

The SFMA approach can strongly influence our understanding and interpretation of the results, particularly for complex real datasets like those presented here: GDP vs life expectancy (LE), physician density vs universal health coverage (UHC), and nurse density vs health coverage. For these datasets, SFMA obtains helpful results, while a cross-section of tools available across Stata, R, and Python fall short: 
StoNED fails to converge, SFA and DEA return results that run counter to conventional wisdom in the field, as described below.   
\begin{itemize}
\item For LE vs. GDP, DEA shows no change in LE for different levels of GDP, while the SFA `frontier' stays far below the upper bound of the data, and reports a very steep (almost vertical) increase in life expectancy at low levels of GDP per capita. The SFMA analysis suggests that LE increases gradually with GDP, albeit more at lowest levels of GDP than higher. That result is made possible due to the outlier-removal functionality available to SFMA. 
\item For health coverage, the DEA analysis suggests that physicians have no effect on UHC above 10 physicians per 10,000 population, a clearly falsifiable conclusion. The SFA and SMFA analyses more plausibly show some benefit of more physicians; SFA gain fails to reach the upper limits of the data.  \end{itemize}
In the sections below, we summarize existing solutions, their limitations, discuss the contributions of SFMA, and present a roadmap for the paper.

\subsection{Stochastic Frontier Analysis (SFA)}

SFA~\cite{aigner1977formulation} was a groundbreaking approach to reformulate deterministic estimation of the frontier production function into a statistical model with stochastic variation. From a statistical perspective, SFA is a mixed effects regression that includes both symmetric random effects as well as asymmetric inefficiencies for the units of analysis (i.e. `firms'). 
The frontier function typically lies above most of the observed outputs for each firm, and is estimated using a maximum likelihood formulation for the observed data assuming both one-sided inefficiencies and random error. The frontier estimates the best output that a firm can hope to achieve given input, while fitted values of the inefficiencies for each firm estimate the improvable gaps in outputs given inputs. Absent observation error, the frontier majorizes all observations. 

SFA analysis has been a topic of considerable interest, with a focus on inefficiency modeling. 
The seminal paper~\cite{aigner1977formulation} used  half-normal and exponential random variables for inefficiencies,  developed an iterative approach to fit the likelihood, and noted the sensitivity of the approach to outliers. Closed form solutions for inefficiencies were provided by~\cite{jondrow1982estimation}. \cite{greene1990gamma} used the gamma distribution for inefficiencies, while \cite{feng2019kde} and \cite{griffin2004dirichlet} modeled inefficiencies using non-parametric kernel density and dirichlet processes, respectively. \cite{battese1995model} allow the inefficiencies to be a function of firm-specific variables and time. 
The Student's T-Half Normal \cite{wheat2019studentt}, the truncated normal \cite{baten2006bangladesh}, and the truncated skewed Laplace \cite{nguyen2014laplace} distributions have also been used to model the inefficiencies. 
 \cite{greene2005fixed} use simulation to compute MLE estimates in cases where a closed form solution of the likelihood are not available.  
 the Student's T-Half Normal \cite{wheat2019studentt}, the truncated normal \cite{baten2006bangladesh}, and the truncated skewed Laplace \cite{nguyen2014laplace} distributions have also been used to model the inefficiencies. 

\subsection{Data Envelopment Analysis and StoNED}

Data envelopment analysis (DEA) uses linear programming to majorize input-output observations from all the firms~\cite{bogetoft2010benchmarking}. The approach is simple and robust, resulting in thousands of cited applications. The downside is that there is no way to account for error or uncertainty with DEA, setting up obvious limitations in many applications and motivating stochastic extensions. 

Stochastic Non-smooth Envelopment of Data (StoNED) is a semi-parametric stochastic method that attemps to address the limitations of DEA to account for noise and uncertainty. StoNED is implemented using two stages: (1)  
applying Convex Nonparametric Least Squares (CNLS) to get a monotonically increasing concave shape that best fits the data~\cite{kuosmanen2012stochastic}, and (2) estimating
conditional expected values using the residuals from the CNLS. In the second stage, inefficiencies and variances are obtained using the Method of Moments (MOM) and Pseudo-likelihood methods (QLE)~\cite{kuosmanen2012stochastic} under specific distributional assumptions.

\subsection{Limitations}

Current SFA approaches require the modeler to supply nonlinear transforms for inputs and outputs so that the frontier can be framed as a linear model. Finding these transforms can be a challenge, and in fact in our numerical experiments with real data we found that even when a natural transform exists (such as log-log), it still very useful to fit a nonparametric frontier to the transformed dataset. Second, statistical models for inefficiencies cannot handle outliers -  the entire frontier can be disrupted by one or a few points that deviate from the general pattern. This requires modelers to pre-process the data, even more than for classic analyses such as linear regression. Finally, existing numerical schemes struggle to converge for large datasets, due to the numerical ill-conditioning of the SFA likelihood function, even in the classic setting where inefficiencies are modeled as half-normal or exponential. SFA requires working with the log of the complementary error function, posing numerical challenges that have gone unaddressed until now.  

The DEA approach is more robust, casting the frontier estimation problem as a linear program, and obtaining a piecewise-linear frontier. Conversely, this approach is unable to account for noise in the data, motivating stochastic extensions, such as StoNED, which provides significant modeling flexiblity. However, because StoNED uses a two-stage approach, the shape of the frontier is fixed before distribution of the errors is accounted for, leaving only the height of the frontier to be determined by residual analysis. In particular, we cannot incorporate any statistics into the fitting process, since they are unknown during the first stage. Despite the focus on stability in the envelopment literature, we encountered difficulties using  StoNED for the real datasets of interest in the global health setting, as discussed in our experiments.  

Finally, neither SFA, nor DEA or StoNED are able to identify and remove outliers using likelihood based methods.

\subsection{Contributions}

Here we develop a non-parametric extension to SFA that also optionally incorporates reported standard errors, a common situation in meta-analysis. We dub the method Stochastic Frontier Meta-Analysis (SFMA). 
We model the envelope as a spline and estimate all coefficients and statistics in an all-at-once likelihood-based approach. The  interface allows the user to incorporate reported standard errors to encode known sampling error, as well as allow for unknown inefficiencies (positive random effects) and non-sampling error (symmetric random effects). The method can be applied to general situations, including population-level problems in global health. 

The functionality of SFMA is compared to other methods in Table~\ref{tab:model_comparison}. 
\begin{table}[htbp]
    \centering
    \caption{Model comparison}
    \label{tab:model_comparison}
    \begin{tabular}{lcccc}
        \hline
        Model & Stochastic & Non-parametric & Specified Obs. SE & Outlier-Robust \\
        \hline
        DEA &  & \checkmark & & \\
        SFA & \checkmark &  & & \\
        StoNED & \checkmark & \checkmark & & \\
        SFMA & \checkmark & \checkmark & \checkmark & \checkmark \\
        \hline
    \end{tabular}
\end{table}

\paragraph{Non-parametric frontier}
We allow the frontier to be modeled using a spline. Scientists using the stochastic frontier may have prior knowledge about the shape of the frontier in relation to a set of inputs. While parametric models (e.g. linear or log-linear) impose very strong assumptions, more typical prior knowledge concerns the shape of the frontier (e.g. monotonically increasing or concave).  We use basis splines to capture nonlinear frontiers with optional constraints on shape of the frontier. The main contrast to StoNED is that the shape is estimated simultaneously with statistics of errors and inefficiencies using a maximum likelihood approach. 

\paragraph{Relative errors of input points}
An analysis at the population level aggregates over input points with reported specific errors, typically  related to the sample sizes of input studies. This  information helps to distinguish more trustworthy points from larger studies/locations from noisier points resulting from smaller sample sizes. Meta-analysis uses these errors to test for the presence of non-sampling error in the data, and to improve the accuracy of the aggregate. 
There is no way to incorpprate reported uncertainty into SFA, DEA, or StoNED, but we can directly use this information through the SFMA interface.

\paragraph{Robust Frontier Estimation Using Trimming}
Real-world datasets often contain outlying firms. SFA, DEA, and StoNED do not offer functionality for automatic detection of outliers. We develop a robust extension of the least trimmed squares (LTS) method~\cite{rousseeuw1984least, aravkin2020trimmed,zheng2021trimmed} for SFA that makes it possible to estimate frontiers for datasets that may contain outliers.

\paragraph{Customized Optimization Algorithm} 
We analyze the likelihood function required to fit SFA and SFMA, and highlight a partial convex structure. We exploit this structure to develop a customized block-coordinate descent algorithm, in effect separating a general nonconvex ill-conditioned problem into a convex problem and a simple scalar nonconvex problem.  The specialized algorithm converges quickly on difficult problems, whereas naive approaches using available tools (e.g. \verb{scipy.minimize{)  fail to solve the SFA, or even evaluate the relevant likelihood function, on simple examples. 

\subsection{Roadmap}
The paper proceeds as follows. In Section~\ref{sec:Setup} we discuss the statistical model for SFMA, derive the likelihood, and discuss basis-splines with shape constraints. We also develop the trimming concept to robustify SFMA. In Section~\ref{sec:Analysis} we analyze the maximum likelihood objectives, present a result on partial convexity, and develop an algorithm for solving the original problem and the trimmed extension. Numerical results using simulated data are presented in Section~\ref{sec:simNumerics}, including comparisons of SFMA to alternative approaches on standard datasets and datasets containing outliers. Applications to real datasets are presented in Section~\ref{sec:realNumerics}. All algorithms are implemented and released in the \verb{sfma{ Python package\footnote{https://github.com/ihmeuw-msca/sfma}.


\section{SFMA Formulation and Likelihoods}
\label{sec:Setup}

In this section, we develop the SFMA model that captures classic SFA, as well as meta-analysis and meta-analytic extensions of SFA analysis. 

\subsection{Statistical Model}
\begin{figure}[h!]
\begin{center}
\includegraphics[scale=0.35]{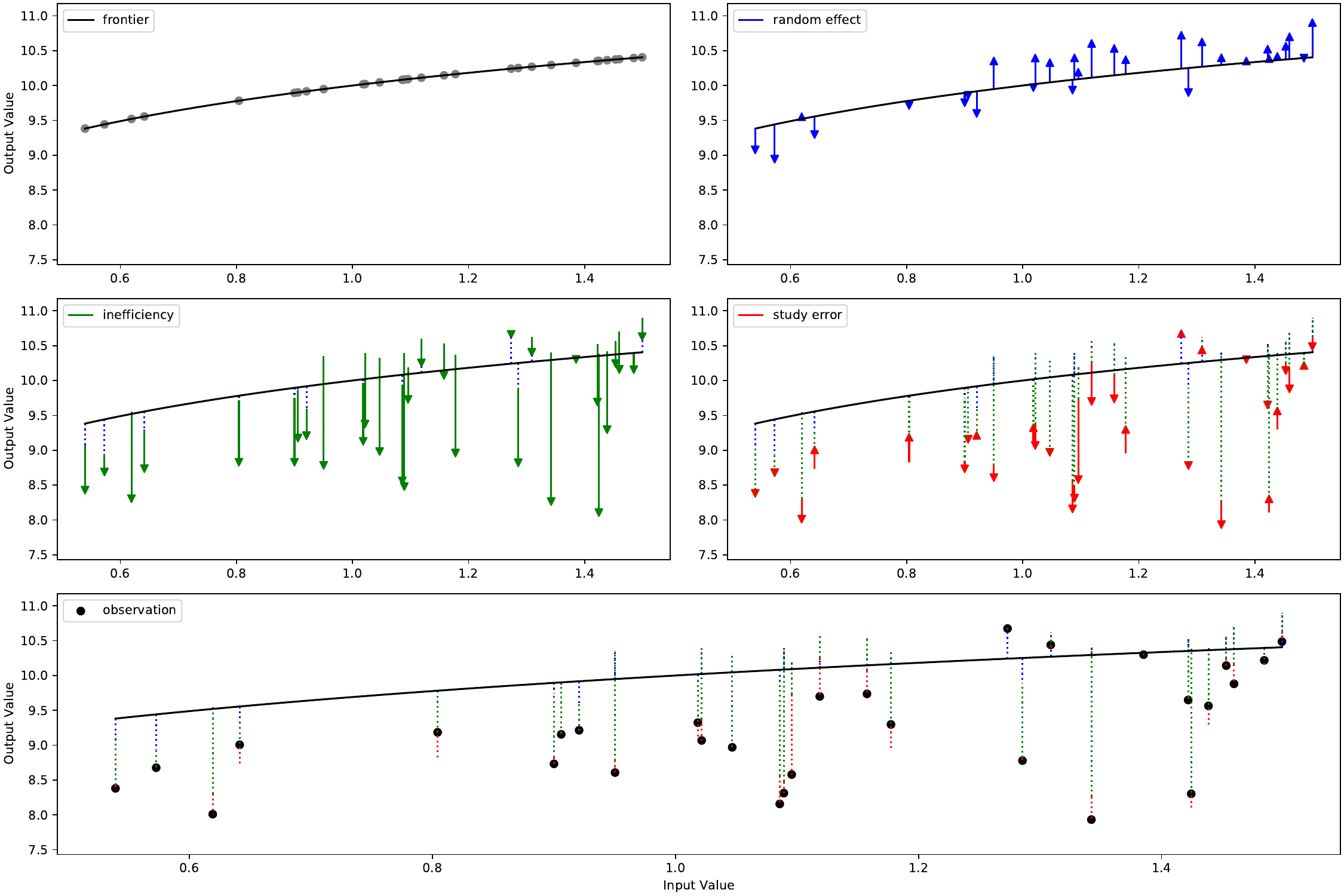}
\caption{Data generating process for stochastic frontier meta-analysis: deviations from the frontier due to random effects, inefficiencies, and study errors. Observed data points are shown in black at the bottom. Generated with half-normal inefficiencies with $\eta = 1$, $\gamma = 0.25$, study-specific errors drawn uniformly between 0 and 0.75, one covariate drawn uniformly between 0.5 to 1.5, and the frontier function $f(x) = \log(x) + 10$.} \label{data-fig}
\end{center}
\end{figure}

The SFMA model is specified as follows: 
\begin{equation}
\label{eq:mr1}
\begin{aligned}
y_i &= \ip{x_i, \beta} +  u_i - v_i  + \epsilon_i \\
\epsilon_i & \sim N(0, \sigma_i^2), & \sigma_i^2 \mbox{ known } \\
u_i & \sim N(0, \gamma) &  \quad \gamma \mbox{ unknown } \\
v_i & \sim HN(0, \eta) &  \quad \eta \mbox{ unknown }
\end{aligned}
\end{equation}
where $y_i$ is the output of firm $i$, $\ip{x_i, \beta}$ encodes the frontier using firm-specific covariates, $\epsilon_i$ are sampling errors with known variances $\sigma^2_i$,
$u_i$ are {\it random effects}, that is, non-sampling errors with unknown variance $\gamma$, and $v_i$ are inefficiencies with  unknown variance $\eta$. The data-generating mechanism for model~\eqref{eq:mr1} is illustrated in Figure~\ref{data-fig}. 
\begin{table}[h!]
\caption{\label{tab:models} Model specifications using~\eqref{eq:mr1}}
    \centering
    \begin{tabular}{c|c|c}\\
Parameter Settings         &  Model type & References  \\\hline 
\{$\sigma_i^2 = 0$\}, $\gamma$ free, $\eta$ free & Classic SFA 
         &  \cite{aigner1977formulation}  \\ \hline 
\{$\sigma_i^2$\} given, $\gamma$ free, $\eta=0$ & Meta-analysis
         &  \cite{borenstein2021introduction, zheng2021trimmed}   \\ \hline 
\{$\sigma_i^2$\} given, $\gamma=0$,  $\eta$ free  & Meta-analytic frontier  
         &  This paper  \\ \hline 
\{$\sigma_i^2$\} given, $\gamma$ free, $\eta$ free & Meta-analytic frontier \\ & with non-sampling error  
         &  This paper  \\ \hline 
    \end{tabular}
\end{table}

Model~\eqref{eq:mr1} captures traditional forms of the SFA, as well as meta-analytic extensions we introduce for population-level analysis, see Table~\ref{tab:models}. We proceed now with developing the approach for~\eqref{eq:mr1}, for simplicity restricting inefficiencies $v_i$ to be half-Normal.

\subsection{A Non-Parametric Frontier Using B-Splines}
\label{sec:spline}

We use a flexible non-parametric spline representation rather than 
pre-specified functional forms of the frontier.
This approach  allows us to maintain flexibility and still incorporate {\it a priori} knowledge about the shape of the frontier. Common assumptions about the frontier function include monotonicity (more input results in larger output) and concavity (diminishing returns with increasing inputs), and are encoded into the spline model using linear constraints as discussed below. 

Basis splines (B-splines) give the nonlinear frontier function a simple representation as a linear combination of basis elements. 
Every spline specification has at least two knots, defining the range of the domain. Using degree $p$ polynomials and $k$ knots at locations by $t_0, ..., t_k$, 
we can represent any frontier as the linear combination of $p+k$ spline basis elements, with coefficients $\beta \in \mathbb{R}^{p+k}$ (see appendix section \ref{sec:appendix-splines} for details):
\begin{equation}\label{eq:spline}
f_{\beta}(t) = \sum_{j=1}^{p+k} \beta_j s_j (t).
\end{equation}
Equation~\eqref{eq:spline} represents arbitrary functions as linear combinations of the basis elements, and is equivalent to equation~\eqref{eq:mr1}  with $x_i$ rows from a spline design matrix, 
see~\eqref{eq:splineDesign} in section \ref{sec:appendix-splines}.


We can specify the shape of the frontier using linear constraints to enforce concavity, convexity, or monotonicity of the spline. These constraints are imposed using derivatives: 
monotonicity is encoded using first derivatives, while convexity/concavity is captured by second derivatives. 
For a basis spline,  derivatives of all orders are linear functions of spline coefficients, and so all constraints can be imposed through the simple representation $C\beta \leq c$ over a fixed grid,
where $C\beta$ returns values of the derivatives on the grid,  and $c$ is the vector specifying required bounds for these derivatives to achieve the shape constraints.

\subsection{Likelihood}
The likelihood corresponding to~\eqref{eq:mr1} is given by 
\begin{equation}
\label{eq:joint}
\begin{aligned}
    p(\beta, &\gamma, \eta, u, v | y) 
        = p(\beta, \gamma, \eta | u, v, y) \cdot p(u, v | y) \\
        &\propto \prod_{i=1}^{n}
            \frac{1}{\sqrt{\sigma^2_i \gamma}}
            \exp\lt(-\frac{(y_i - \ip{x_i, \beta} - u_i + v_i)^2}{2\sigma_i^2}
            -\frac{u_i^2}{2\gamma}\rt)
            f(v_i|\eta)\\
        &:= \prod_{i=1}^n \mathcal{L}_i(\beta, \gamma, \eta, u, v).
\end{aligned}
\end{equation}

The model for $v_i$ is left unspecified in~\eqref{eq:joint}. The half-normal model 
\begin{equation}
\label{eq:half-normal}
    f(v_i | \eta) = 
        \begin{cases} 
            \frac{\sqrt{2}}{\sqrt{\pi\eta}} \exp\left(-\frac{v_i^2}{2\eta}\right) & v_i \geq 0 \\
            0 & v_i < 0
        \end{cases}
\end{equation}
is widely used in practice~\cite{baten2006bangladesh},\cite{musau2021norway},\cite{raghunathan2011mfi},\cite{guerrini2018italy}. We stay with this choice throughout the paper, and refer the reader to other models  in the literature, including Gamma \cite{greene1990gamma}, the Student's T-Half Normal \cite{wheat2019studentt}, the truncated normal \cite{baten2006bangladesh}, and the truncated skewed Laplace \cite{nguyen2014laplace} distributions. 

When $v_i$ are half-normal~\eqref{eq:half-normal}, likelihood $p(\beta, \gamma, \eta, u, v | y)$ in~\eqref{eq:joint} is given by  
\begin{equation}
\label{eq:jointOptHN}
\begin{aligned}
     &\prod_{i=1}^{n}
            \frac{1}{\sqrt{\sigma^2_i \gamma\eta}}
            \exp\lt(-\frac{(y_i - \ip{x_i, \beta} - u_i + v_i)^2}{2\sigma_i^2}
            -\frac{u_i^2}{2\gamma}
            -\frac{v_i^2}{2\eta}
            \rt)
\end{aligned}
\end{equation}
with support for each $v_i$ on the positive real line. 
Integrating out the random effects, and taking the negative log of the resulting distribution, we get a closed form solution of the marginal likelihood that only depends on $(\beta, \eta, \gamma)$. 
To write this likelihood compactly, we first define $\erfc$ to be the complementary error function
\begin{equation}
    \label{eq:erfc}
\erfc(z) = 1- \frac{2}{\sqrt{\pi}}\int_{0}^z \exp(-t^2)dt.
\end{equation}
and also define 
\[
 r_i(\beta) = y_i - \ip{x_i, \beta}, \quad V_i(\gamma,\eta) = (\gamma +\eta +  \sigma_i^2).
\]
Integrating $v_i$ and $u_i$ out of~\eqref{eq:jointOptHN}, and taking the negative logarithm, we obtain the negative log likelihood used to fit $(\beta, \eta, \gamma)$:
\begin{equation}
\label{eq:meta_simple_hn}
\begin{aligned}
\ell_{hn}( \beta, \gamma, \eta| y)
&= -\ln\lt(\int_{\mathbb{R}^m_+}\int_{\mathbb{R}^m} p(\beta, \gamma, \eta,  u,  v | y)\, du dv\rt) \\
=& \sum_{i=1}^m \frac{r_i(\beta)^2}{2V_i(\gamma,\eta)} + \frac{1}{2}\ln V_i
 - \ln\left(\erfc\left(\frac{\sqrt{\eta}r_i(\beta)}{\sqrt{2(\gamma + \sigma_i^2) V_i(\gamma,\eta) }}\right)\right)\\
\end{aligned}
\end{equation}
The $\ln\erfc$ function in~\eqref{eq:meta_simple_hn} presents a technical challenge to algorithmic development, and is discussed in Section~\ref{sec:Analysis} along with other  details.

The SFMA model is fit by minimizing~\eqref{eq:meta_simple_hn} subject to linear inequality constraints that capture features such as monotonicity and convexity/concavity (see  Section~\ref{sec:spline}) using the algorithm described in Section~\ref{sec:blockcoord}. The estimated coefficients $\hat \beta$ give us the estimated frontier and any other covariate multipliers of interest, while estimated variances $(\hat \gamma, \hat \eta)$ are used to estimate individual non-sampling errors $\hat u_i$ and inefficiences $\hat v_i$ as described in Section~\ref{sec:firms}. 
 In the next section, we discuss the trimming concept that modifies~\eqref{eq:meta_simple_hn} to make the proposed SFMA approach robust to outliers.

\subsection{Extensions for Trimming}
\label{sec:Trimming}

Outliers make it difficult to estimate a frontier using SFA~\cite{aigner1977formulation}. 
For example, a large outlier at low input can easily skew the entire frontier. 
\textit{Trimming estimators} from robust statistics~\cite{rousseeuw2005robust} seek to simulatenously estimate a model and automatically discover such outliers. Though trimming has a long history (see e.g.~\cite{rousseeuw1984least}), 
it has recently been extended to general learning models  
~\cite{AravkinDavis2020} and to meta-analysis~\cite{zheng2021trimmed}. Here we extend the trimming concept to the SFMA likelihood in order to get an outlier-robust benchmarking method.  

Writing the negative log likelihood problem in~\eqref{eq:meta_simple_hn} as a sum over individual contributions over the dataset
\begin{equation}
\label{eq:sumlik}
    \min_{\theta} \sum_{i=1}^n \ell_i(\theta) \quad\mbox{s.t.}\quad   C\theta \leq c
\end{equation}
with $\theta := (\beta, \gamma, \eta)$, we now considering auxiliary weights $w_i$, one per datapoint, with constraints that each weight is in the interval $[0,1]$ and all the weights sum up to the specified number of inliers: 
\begin{equation}\label{eq:trim}
\arg\min_{\theta, w} \sum_{i=1}^n w_i \ell_i(\theta), \quad \mbox{s.t.} \quad  C\theta \leq c, \quad  0 \leq w_i \leq 1, \quad 1^T w = h.  
\end{equation}
If $h = n$, then problem~\eqref{eq:sumlik} requires that every $w_i = 1$ and we reduce to the original SFMA likelihood. For $h < n$, the weights are preferentially placed on the $h$ ``best fitting'' observations (inliers), as judged by the likelihood at a given $\theta$; the likelihood is in turn is affected by which points are outliered in this manner. Since we  optimize over both $\theta$ and $w$,  the joint problem is highly nonconvex~\cite{AravkinDavis2019}, but both theoretical and empirical results show the approach is stable and useful. Automated strategies for choosing $h$ is an open problem, and in practice the analyst chooses a common level across datasets, e.g. fitting 90\% of the data to hedge the quality of the results against potential outliers while still getting information from the bulk of the data~\cite{murray2020global}. The method we use to estimate~\eqref{eq:sumlik} leverages the developments in~\cite{zheng2021trimmed} for trimmed meta-analysis, and is discussed in Section~\ref{sec:trimalgo}.

\section{Optimization Methods and Analysis}
\label{sec:Analysis}

In this section we describe the optimization methods used to solve the SFMA likelihood~\eqref{eq:jointOptHN} as well as the corresponding trimmed extension~\eqref{eq:trim}. In order to do this, we discuss technical challenges and theoretical properties of the $\ln\erfc$ function, which plays an outsized role in~\eqref{eq:jointOptHN}. 
This section proceeds as follows. In Section~\ref{sec:lnerf} we discuss $\ln\erfc$ characterizes its convexity. This leads to a customized block-coordinate descent method for~\eqref{eq:jointOptHN} in Section~\ref{sec:blockcoord}, 
where the problem is decomposed into an ill-conditioned convex problem with constraints (solved using an interior point method)  and two simple scalar nonconvex problems that are easily solved. 
Finally, the algorithm for solving the trimmed likelihood~\eqref{eq:trim} discussed in Section~\ref{sec:trimalgo}. 

\subsection{The log of the complement to the Gaussian Error}
\label{sec:lnerf}

The $\ln\erfc$ function
\[
\ln\erfc(z) = \log\left(1- \frac{2}{\sqrt{2\pi}}\int_0^z \exp(-t^2)dt\right)
\]
plays a central role in the likelihood formulation and analysis for the SFMA application, see~\eqref{eq:jointOptHN}. Simply evaluating the $\ln\erfc$ function  presents technical challenges, and current packages in Python do not evaluate it correctly: there are no standard implementations of $\ln\erfc$ in \texttt{numpy}/\texttt{scipy} in Python or in base R, and specialized routines in Python and R return wrong values for large inputs.
For example, the function \texttt{erfc} in \texttt{scipy.special.erfc} and the \texttt{erfc} function in the \texttt{pracma} package in R will both return exactly 0 for $x \geq 27$, 
and $\ln\erfc(x)$ will correspondingly return $-\infty$ for these values in both implementations. This is likely because inputs in this range are not meaningful for simple statistical tests, but the issue plays a major role in an optimization setting, where it causes optimization techniques to fail for even moderately sized problems.  

In order to fix the implementation of the basic function, we 
used the asymptotic expansion for large $x$ and then created our own stable implementation of $\ln\erfc$ that uses the asymptotic expansion for large values. The asymptotic expansion of $\erfc(x)$ is given by~\cite{hunter2004asymptotic}:
\begin{equation}
\label{eq:erfc_expansion}
\erfc(x) = \frac{\exp(-x^2)}{\sqrt{\pi}x}\left(
1 + \sum_{n=1}^\infty(-1)^n\frac{(2n - 1)!!}{(2x^2)^n}
\right)
\end{equation}
We use the first two terms to approximate $\erfc(x)$ for large $x$:
\begin{equation}
    \label{eq:erfc_approx}
    \erfc(x) \approx \frac{\exp(-x^2)}{\sqrt{\pi}x}\left(
    1 - \frac{1}{2x^2}
    \right)
\end{equation}
And for $\ln\erfc(x)$, we have
\begin{equation}
    \label{eq:log_erfc_approx}
    \ln(\erfc(x)) \approx -x^2 + \ln\left(1 - \frac{1}{2x^2}\right) - \ln(\sqrt{\pi}x)
\end{equation}
For our Python implementation, we use standard routines when $x < 25$ and \eqref{eq:erfc_approx} for $x\geq 25$. This yields a stable implementation for the objective function and makes it possible to optimize the required likelihoods. 

Here, we establish concavity of $\ln\erfc$ and its approximation~\eqref{eq:log_erfc_approx}. This allows us to show that the likelihood~\eqref{eq:jointOptHN} is convex with respect to $\beta$. 

\begin{theorem}[Concavity of $\ln\erfc$ and its approximation]
\label{thm:h_concave}
The function $\ln\erfc$ is concave, and so is its approximation~\eqref{eq:log_erfc_approx} for $x\geq 25$.
\end{theorem}
The proof is given in the Appendix.

\begin{corollary}[Partial convexity]
\label{cor:convex}
The marginal likelihood~\eqref{eq:meta_simple_hn} is convex with respect to $\beta$. 
\end{corollary}

\begin{proof}

The likelihood~\eqref{eq:meta_simple_hn}  depend on $\beta$ through compositions of affine\footnote{Affine functions are linear functions with an offset, e.g. $f(x) = Ax-b$. Compositions of affine functions with convex functions are convex.} functions $r_i(\beta)$ with convex functions: affine, quadratic, and $-\ln\erfc$. 
\end{proof}

In the next section, we describe the block-coordinate descent approach that takes advantage of the partial convexity structure in Corollary~\ref{cor:convex}.

\subsection{Block-coordinate Descent}
\label{sec:blockcoord}

The partial convexity of~\eqref{eq:jointOptHN} with respect to $\beta$ makes it possible to solve the problem using a block-coordinate descent algorithm, that alternates full solves in blocks $(\beta, \gamma, \eta)$ until convergence. 
The largest block is $\beta$, which may contain multiple parameters, including spline coefficients and covariate multipliers, as well as constraints $C\beta \leq c$. Convexity of the objective with respect to $\beta$ makes it possible to optimize over $\beta$ efficiently using an interior point method described in Section~\ref{sec:interior} of the Appendix. 
The remaining blocks are scalars, and we easily solve these problem using root finding.  The convergence of the block-coordinate scheme here follows from classic results. 

\begin{theorem}[Convergence of Block-Coordinate Descent for SFA Marginal Likelihood]

Suppose that the solution of~\eqref{eq:jointOptHN} with respect to $\beta$ is unique for each value of $\eta$ and $\gamma$, e.g. if the design matrix for the problem $X^TX$ has full rank. Then the block coordinate scheme
\[
\begin{aligned}
\beta^{k+1} &= \arg\min_\beta \ell(\beta, \gamma^k, \eta^k) \\
\gamma^{k+1} &= \arg\min_\gamma \ell(\beta^{k+1}, \gamma, \eta) \\
\eta^{k+1} & = \arg\min_{\eta} \ell(\beta^{k+1}, \gamma^{k+1}, \eta)
\end{aligned}
\]
converges to a stationary point of~\eqref{eq:jointOptHN}. 
\end{theorem}

\begin{proof}
The convergence scheme satisfies conditions required by~\cite{tseng2001convergence}[Theorem 4.1 case (c)]. Since there are only three blocks, as long as the convex problem in $\beta$ has a unique solution, we can apply the theorem. 
\end{proof}

In the next section, we show how to solve the extended likelihood~\eqref{eq:trim}, leveraging the block-coordinate descent algorithm in this section.

\subsection{Trimming Method}
\label{sec:trimalgo}
The algorithm for trimming uses variable projection~\cite{bell1996relative,golub2003separable,aravkin2012estimating}. Considering the joint likelihood~\eqref{eq:trim}, 
and let $\theta = (\beta,\gamma, \eta)$.
Define the value function $v(w)$ and optimal value 
$\theta(w)$ by  
\begin{equation}
\label{eq:value}
\begin{aligned}
v(w) &= \min_{\theta} \sum_{i=1}^m w_i \ell_i(\theta) \quad\mbox{s.t.}\quad   C\theta \leq c\\
 \theta(w) & = \arg\min_{\theta}  \sum_{i=1}^m w_i \ell_i(\theta) \quad\mbox{s.t.}\quad   C\theta \leq c.
\end{aligned}
\end{equation}
The term {\it variable projection}  refers to partially minimizing over $\theta$. We use the block-coordinate solver in Section~\ref{sec:blockcoord} to solve the problem for each fixed $w$, reducing the overall problem to value function optimization
\begin{equation}
\label{eq:capped}
\min_{w} v(w) \quad \mbox{s.t.} \quad w \in \Delta_h:=\{w: 0\leq w_i \leq 1, \; 1^Tw = h\}. 
\end{equation}
where $v(w)$ is differentiable with derivative given by~\cite[Theorem 2]{zheng2021trimmed}
\begin{equation}
\label{eq:vgrad}
\nabla v(w) = \nabla_w \lt(\sum_{i=1}^m w_i \ell_i(\theta)\rt)|_{\theta = \theta(w)} =  \begin{bmatrix} \ell_1(\theta(w)) \\ \vdots \\ \ell_m(\theta(w))\end{bmatrix}.
\end{equation}
The top level algorithm is then simply a projected gradient method 
where at each iteration we compute the gradient and project the updated point onto the set $\Delta_h$ defined in~\eqref{eq:capped}: 
\[
 w^+ = \mbox{proj}_{\Delta_h} (w - \alpha \nabla v(w)).
\]
Each evaluation of $\nabla v$ requires a full minimization step over the constrained weighted likelihood 
with respect to $\theta$ using our block-coordinate descent method. The capped simplex $\Delta_h$ is a closed convex set with a simple projection;  proximal gradient with line search converges in this case 
for any positive step-size $\alpha$~\cite{AravkinDavis2019}.

\subsection{Estimating Firm-Specific Inefficiencies}
\label{sec:firms}
Once fixed effects $\theta $ have been estimated, we want to obtain estimates of both heterogeneity and inefficiency from the joint likelihood~\eqref{eq:jointOptHN}. 
Using $r_i$ to denote the residuals
\[
r_i = y_i - \ip{ x_i, \beta},
\]
we optimize
\begin{equation}
\label{eq:REhn}
\begin{aligned}
\min_{u_i, v_i \geq 0} \frac{(r_i - u_i +v_i)^2}{2\sigma_i^2} + 
\frac{u_i^2}{2\gamma}+
\frac{v_i^2}{2\eta}
\end{aligned}
\end{equation}

Partially minimizing with respect to $u_i$, we obtain 
\begin{equation}
\label{uofv}
\begin{aligned}
u_i(v_i) & = \frac{\gamma(r_i  +v_i)}{\sigma_i^2 + \gamma}
\end{aligned}
\end{equation}
which shows that $\hat u_i$ is fully determined by $\hat v_i$. To solve for the latter, we plug the estimate back into the expression of interest, 
obtaining 
\begin{equation}
\label{valuevu}
\min_{u_i}\frac{(r_i - u_i +v_i)^2}{2\sigma_i^2} +
\frac{u_i^2}{2\gamma}
= \frac{1}{2(\sigma_i^2 + \gamma)}(r_i  +v_i)^2.
\end{equation}
This reduces the problem to 
\begin{equation}
\label{eq:REhnRed}
\begin{aligned}
\min_{v_i \geq 0} &\frac{1}{2(\sigma_i^2 + \gamma)}(r_i  +v_i)^2 + \frac{v_i^2}{2\eta}\\
\hat v_i & = \max\left(0, \frac{-\frac{1}{\sigma_i^2 + \gamma}r_i}{\frac{1}{\sigma_i^2 + \gamma} + \frac{1}{\eta}}\right) \\
& = \max\left(0, -\frac{\eta r_i}{\sigma_i^2 + \eta + \gamma}\right)
\end{aligned}
\end{equation}


\section{Numerical Experiments}
\label{sec:Numerics}
In this section we present results for synthetic and real datasets comparing SFMA to competing alternatives. 

\subsection{Synthetic Experiments}
\label{sec:simNumerics}
We created four different simulations and compared the results across SFMA, DEA, StoNED, and SFA packages.

\subsubsection{Simulation specifications}

In all simulations,  $x_i$ denote exposures of the independent variable,        
 $y_i$ denote the output of the firm $i$, $u_i$ is the firm-specific inefficiency with variance $\sigma^2_u$ while $\epsilon_i$ is a random error with standard error $ \sigma^2_\epsilon$. 
For simulations 1-3, we used 
\[
\begin{aligned}
x_i &= U(0,1) \\
y_i &= 3 + \log(x_i + 0.2) + \epsilon_i - u_i \\
u_i &= HN(0,\sigma_u^2), \quad 
\epsilon_i = N(0,\sigma_\epsilon ^2)  \\
\end{aligned}
\]
while for simulation 4, we used 
\[
y_i = 
\begin{cases}
3 + \log(x_i + 0.2) + \epsilon_i - u_i & \text{if inliers} \\
10 + \log(x_i + 0.2) +\epsilon_i - u_i  &\text{if outliers}
\end{cases} 
\]
with parameters as recorded in Table~\ref{tab:sim_spec}. The simulations become increasingly difficult, with Sim 3 containing a mixture of reported standard errors, and Sim 4 also containing outliers.

\begin{table}[h!]
    \centering
    \caption{\label{tab:sim_spec} Simulation Specifications}
    \begin{tabular}{l|c|c|c|c}\hline 
    Sim  & $\sigma_u^2$   & $\sigma_\epsilon^2$ & Outlier $\%$   & Data Points \\ \hline
         1            &  $1$      &  $0.2$       & 
        $0\%$      & 200 \\ \hline
        2            &  $1$      &  $\sqrt{0.2 x_i}$       & 
        $ 0\% $      &   200 \\ \hline
         3           & $0.5$      & 
         \( 
         \begin{aligned} 
         & 0.05 (67\% \text{ of data}) \\ 
         & 1.0 (33\% \text{ of data)} 
         \end{aligned}
         \)
         & 0\%       &  210\\\hline 
     4           & $0.5$      & 
         \( 
         \begin{aligned} 
         & 0.05 (67\% \text{ of data}) \\ 
         & 1.0 (33\% \text{ of data)} 
         \end{aligned}
         \)
         & 12.5\%       &  210\\\hline 
    \end{tabular}
\end{table}

\subsubsection{Results for Simulated data}

\begin{table}[h!]
    \centering
    \caption{\label{tab:rmse_values} RMSE values for different models}
    \begin{tabular}{l|c|c|c|c}\hline 
        Sim  & Sim 1  & Sim 2  & Sim 3  & Sim 4  \\ \hline
        DEA            & 0.017290      & 0.33918       & 1.08809       & 62.54372      \\
        SFA            & 0.05109       & 0.16333       & 0.01679       & 0.40832       \\
        StoNED (QLE)   & 0.00360       & 0.01170       & 0.01551       & 5.81288       \\
        StoNED (MOM)   & 0.00378       & 0.01179       & 0.01399       & 0.31752       \\
        SFMA           & {\bf 0.00223}       & {\bf 0.00399}       & {\bf 0.00186}       & 0.93045       \\
        R-SFMA         & -             & -             & -             & {\bf 0.00105}       \\\hline
    \end{tabular}
\end{table}

We compared multiple methods across the four simulations described above. For each example, we ran a monte carlo simulations with 200 random realizations. 
In each scenario and realization, we compared DEA, SFA, two versions of StoNED (MOM and QLE), our method without trimming (SFMA), and our method with trimming (R-SFMA). We also compared the frontiers of production functions generated to the true generating function using root mean squared error (RMSE), summarized in Table \ref{tab:rmse_values}. The best-performing model in each sim is highlighted in bold. 

\begin{figure}[h!]
\begin{center}
\includegraphics[scale=0.65]{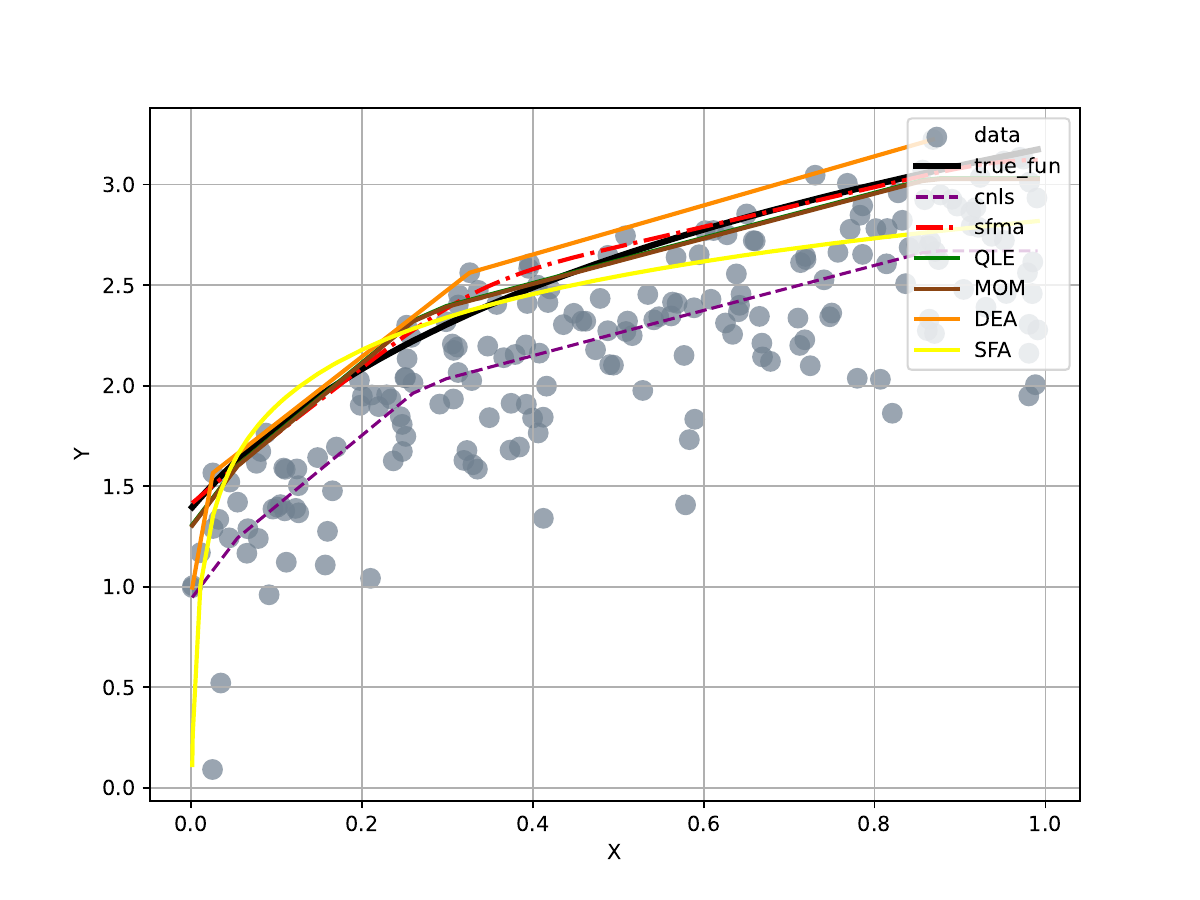}
\caption{Results for Sim 1. SFMA and StoNED methods are close to the true frontier, with SFA below and DEA above. RMSE for results across 200 realizations are in Table~\ref{tab:rmse_values}.} \label{homosk}
\end{center}
\end{figure}

Results for Sim 1 (homoskedastic errors) are shown in Figure  \ref{homosk}. The methods are comparable and reasonably close to the true frontier, with the exception of SFA, which deviates because of choice of functional form. SFMA obtains the best results across multiple random realizations, as summarized in Table~\ref{tab:rmse_values}. 
\begin{figure}[h!]
\begin{center}
\includegraphics[scale=0.65]{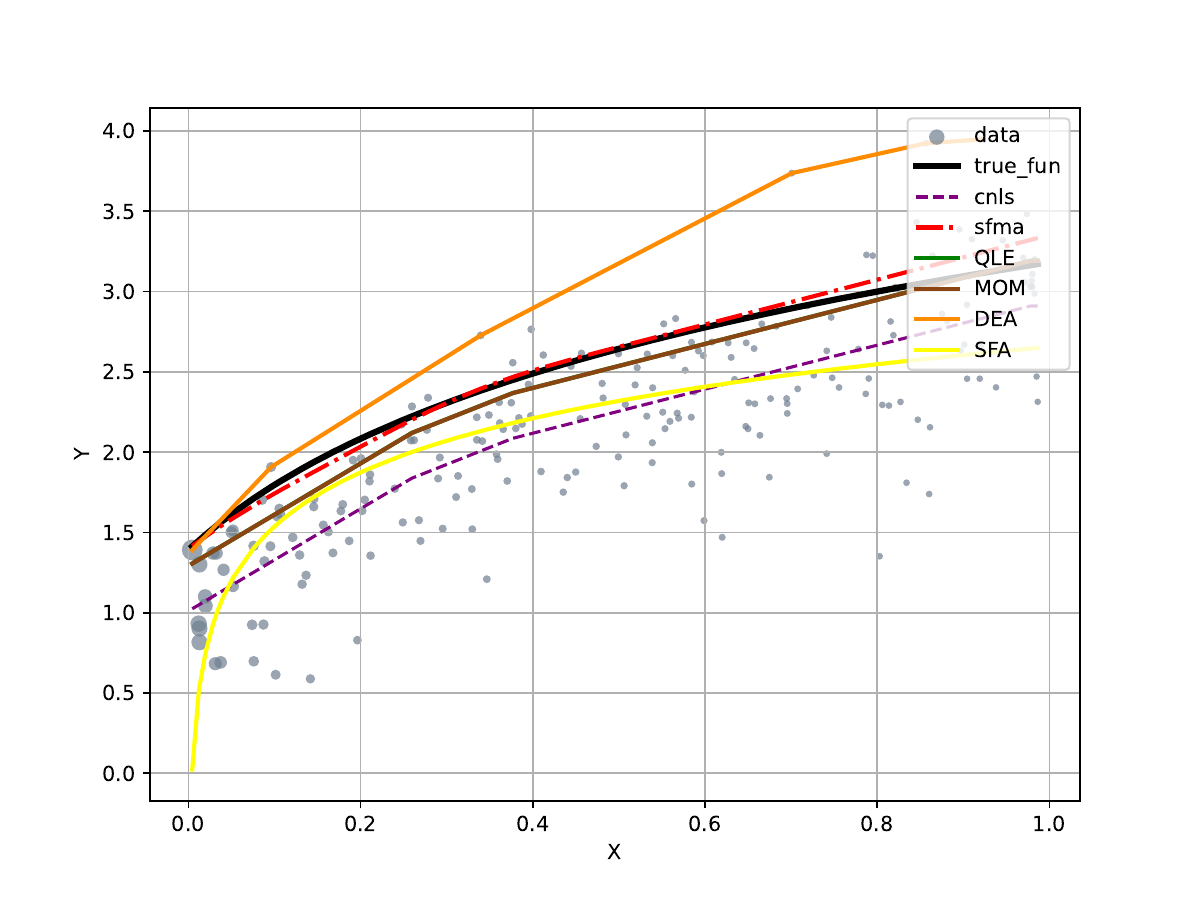}
\caption{Results for Sim 2. SFMA is closest to true frontier, while StoNED methods and SFA are below and DEA is significantly above. RMSE for results across 200 realizations are in Table~\ref{tab:rmse_values}. } \label{heterosk}
\end{center}
\end{figure}
Results for Sim 2 (heteroscedastic errors) are shown in Figure  \ref{heterosk}. Point sizes are inversely proportional to the reported SE of the data points. SFMA has an advantage here, as it directly uses this information in the likelihood. This is reflected in both the realization shown in the figure and in the overall RMSE results in Table~\ref{tab:rmse_values}.

\begin{figure}[h!]
\begin{center}
\includegraphics[scale=0.65]{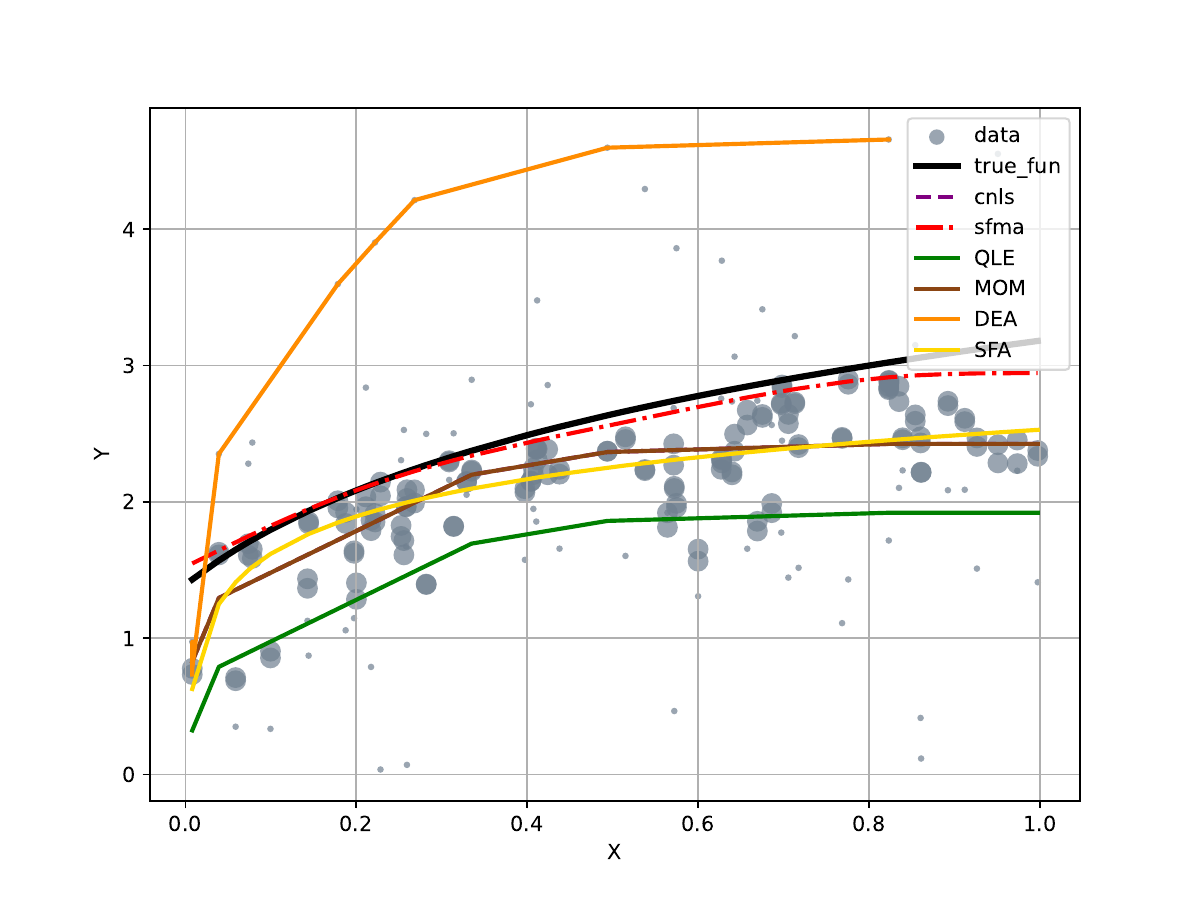}
\caption{Results for Sim 3. SFMA significantly outperforms all other methods as it is able to incorporate relative errors reported for all datapoints. RMSE for results across 200 realizations are in Table~\ref{tab:rmse_values}.} \label{vary_SE}
\end{center}
\end{figure}

Results for Sim 3 (mix of different SE levels) are shown in Figure  \ref{vary_SE}. Point sizes are inversely proportional to the reported SE of the data points. The fact that different SE levels persist across the domain of interest underscores how helpful it is to use this information. SFMA has a significant advantage compared to all the other methods. 

\begin{figure}[h!]
\begin{center}
\includegraphics[scale=0.65]{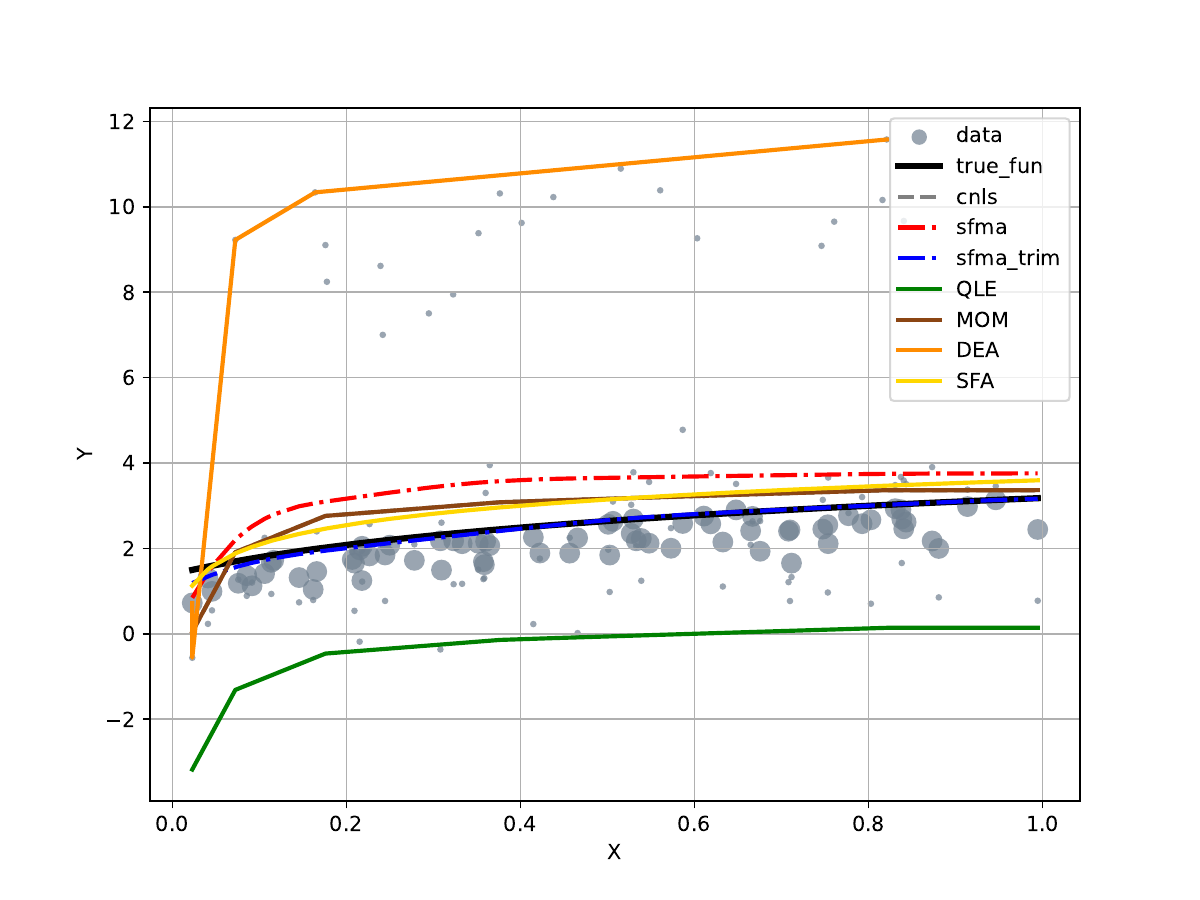}
\caption{Results for Sim 4. SFMA with trimming (dubbed sfma$\_$trim in the plot) recovers the true frontier, while remaining methods deviate from truth because of outliers added to the dataset. RMSE for results across 200 realizations are in Table~\ref{tab:rmse_values}.} \label{outl}
\end{center}
\end{figure}
Results obtained in the presence of outliers are shown in figure \ref{outl}. Here, SFMA with trimming offers a unique capability that is not available in any of the other methods. We also see the potential for huge errors with straightforward envelopment analysis from the large deviation of the DEA result.

\begin{table}[h!]
    \centering
    \caption{\label{tab:param_tab} List of Model Parameters}
    \begin{tabular}{l|c|c|c|c}\hline 
    Sim  & No. knots  & Degree  & Trimming (for SFMA Trim)   & Constraints \\ \hline
        Sim 1            & 7      & 3       & 0      & Increasing, Concave \\
        Sim 2           & 7      & 3       & 0       & Increasing, Concave  \\
        Sim 3   & 7       & 3      & 0       & Inreasing, Concave\\
        Sim 4   & 7       & 3       & 12.5\%  & Increasing, Concave    \\\hline
    \end{tabular}
\end{table}

Table \ref{tab:param_tab} shows the list of parameters used to generate the plots.

\subsection{Real datasets}
\label{sec:realNumerics}
In this section, we consider several challenging datasets used for benchmarking in the global health setting. The first dataset focuses on life expectancy (LE) as a function of GDP, while the second looks at a measure of universal health coverage (UHC) as a function of physicians or nurses per 10,000 people. 

\subsubsection{World bank data}

The World Bank dataset\footnote{https://data.worldbank.org/indicator/SP.DYN.LE00.IN} contains life expectancy from birth as a function of GDP across multiple countries and years around the world. This dataset can be used to benchmark life expectancy across 204 countries over the time period 1960-2021.
A frontier analysis alows us to evaluate the inefficiency for any country-year at 
a given GDP per capita, helping to benchmark how effective countries are in utilizing their GDP to improve life expectancy.  

\begin{figure}[h!]
\begin{center}
\includegraphics[scale=0.4]{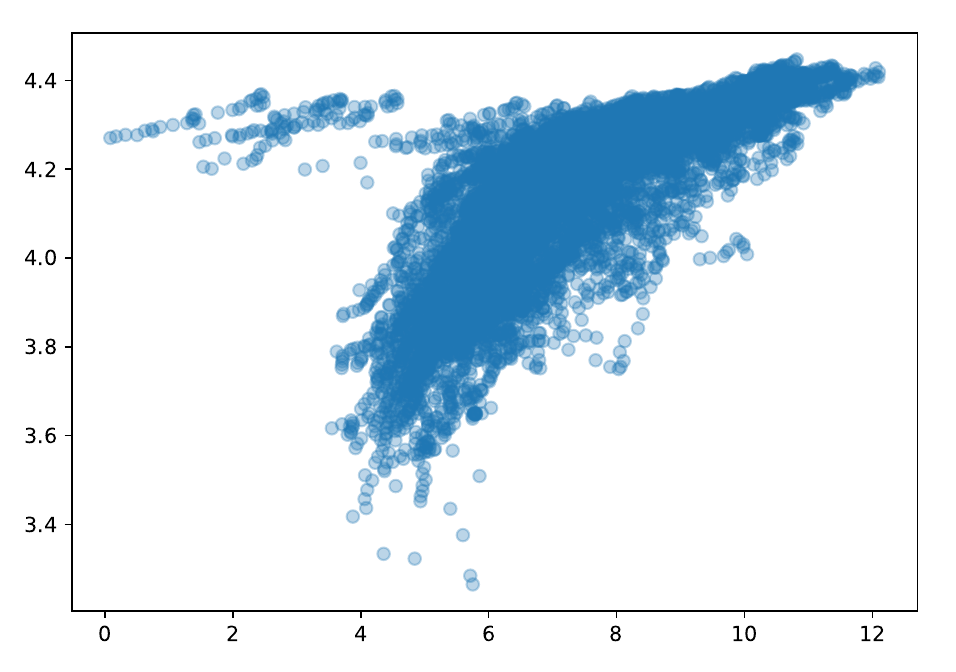}
\includegraphics[scale=0.4]{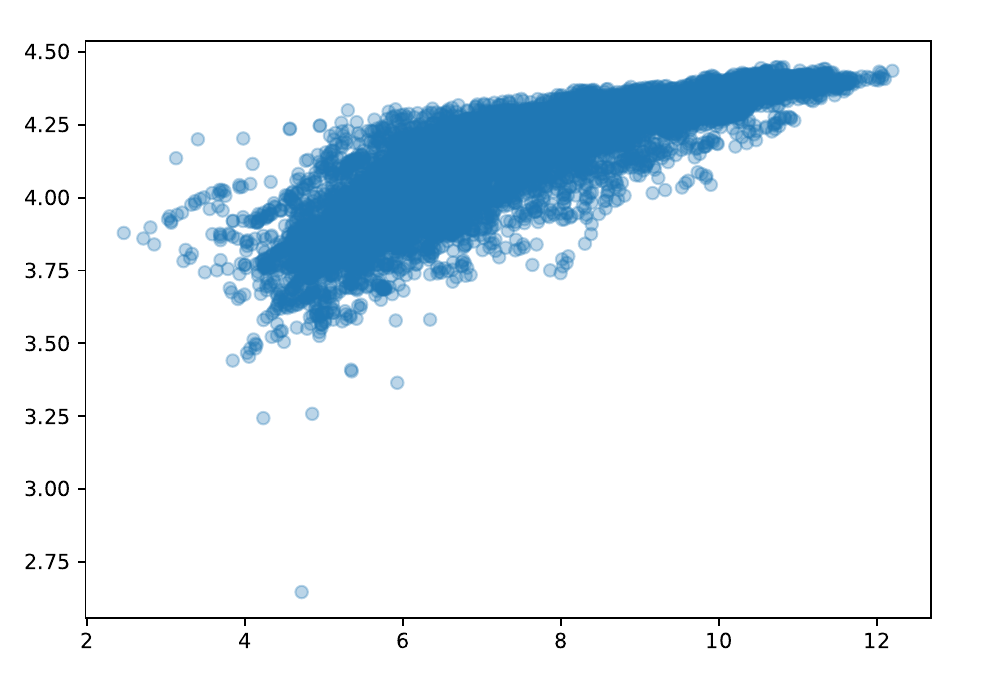}
\caption{Life expectancy as a function of GDP (on log scale). World bank originally posted the data on the left panel, which includes a swath of country-years far away from the bulk of the data. The current version of the data, shown in the right panel, no longer contains the artifacts. We used the left dataset to fit frontiers, to show the outlier-robust functionality of the current method.} \label{gdp_versions}
\end{center}
\end{figure}

In most datasets, `outliers' are a subjective notion. In this case, however, multiple uploads from the World Bank gave us a unique opportunity to confirm a set of outliers that was corrected by the World Bank. In the two uploads shown in Figure~\ref{gdp_versions}, we see a set of country-years with high life expectancy and low GDP (left panel) in the 2021 upload that were later corrected (right panel) in the 2024 upload. We use the left dataset for the analysis to showcase the robust trimming capability of SFMA. 

\begin{figure}[h!]
\begin{center}
\includegraphics[scale=0.65]{figs/gdp_le_log_log.pdf}
\includegraphics[scale=0.65]{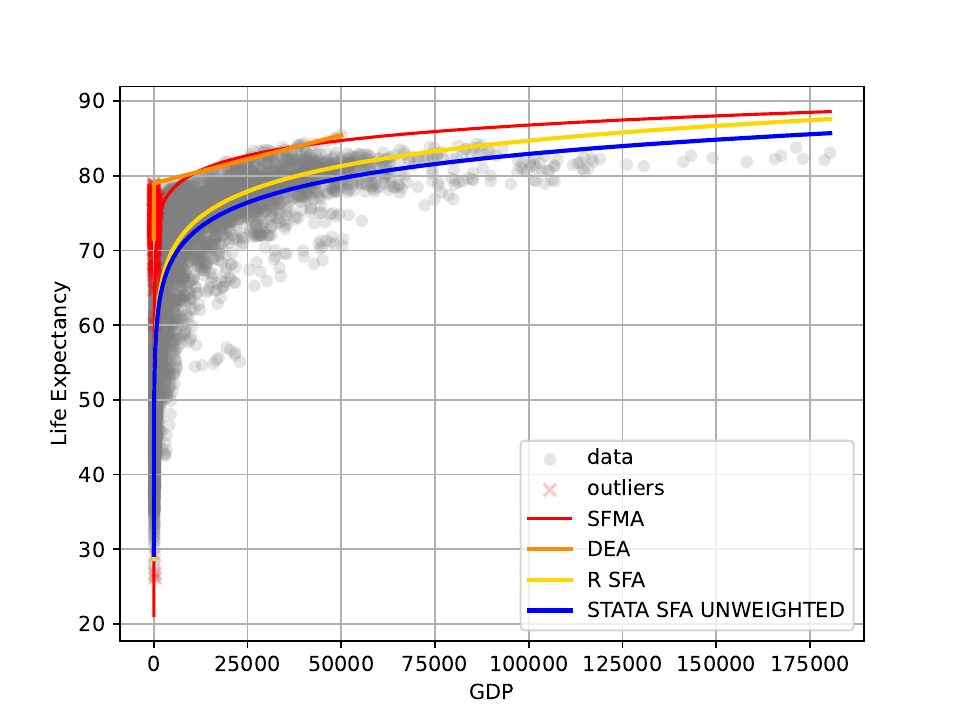}
\caption{Life expectancy as a function of GDP, log scale (top) and linear scale (bottom). DEA creates a flat line across the top, whie R SFA and Stata SFA fit the middle of the data here. SFMA with trimming is able to get a much more useful frontier by ignoring only 3\% of the data. } \label{gdp_le}
\end{center}
\end{figure}

Figure \ref{gdp_le} shows the frontiers built using SFMA with trimming, compared to DEA and SFA. The spline feature of SFMA is very helpful here, since the frontier is clearly nonlinear even after the classic log-log transform. 
The trimming capability is key for this example -- by allowing the model to identify and exclude 3\% of the data as outliers, SFMA gets a very informative fit, staying ahead of the main data cloud while ignoring the spurious data in the top left quadrant. SFA fits a curve essentially through the middle of the data cloud, while DEA stays above both the main cloud and the outliers. The SFA result was so surprising that we also fit the data using the Stata package \verb{sfcross{, and the result is also shown in Figure~\ref{gdp_le}, dubbed `Stata SFA Unweighted'. 
We also attempted to apply the StoNED approach on this example, but the CNLS fit in the first stage of the method failed. 

\subsubsection{UHC data}
The UHC dataset was compiled by IHME\footnote{\url{ghdx.healthdata.org/record/ihme-data/gbd-2019-uhc-effective-coverage-index-1990-2019}}
in order to estimate global progress towards universal health coverage (UHC) in 204 countries and territories in 1990, 2010, and 2019. 
We use this dataset to fit a frontier of the UHC index as a function of Physician density per 10,000 and Nurse density per 10,000, also compiled by IHME\footnote{\url{ghdx.healthdata.org/record/ihme-data/gbd-2019-human-resources-health-1990-2019}}. 

\begin{figure}[h!]
\begin{center}
\includegraphics[scale=0.55]{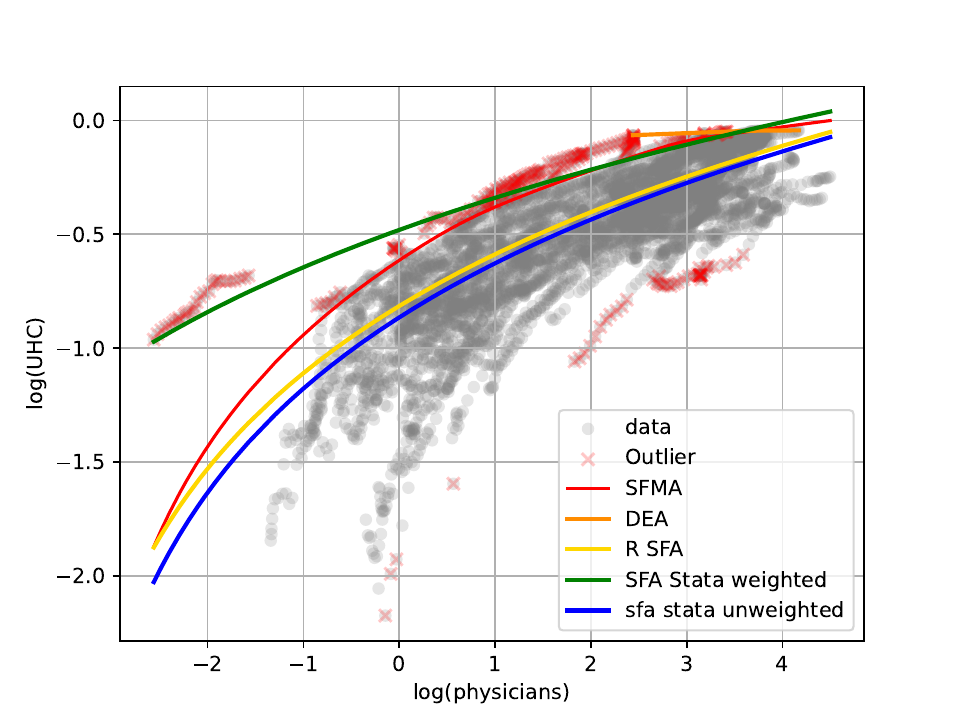}
\includegraphics[scale=0.55]{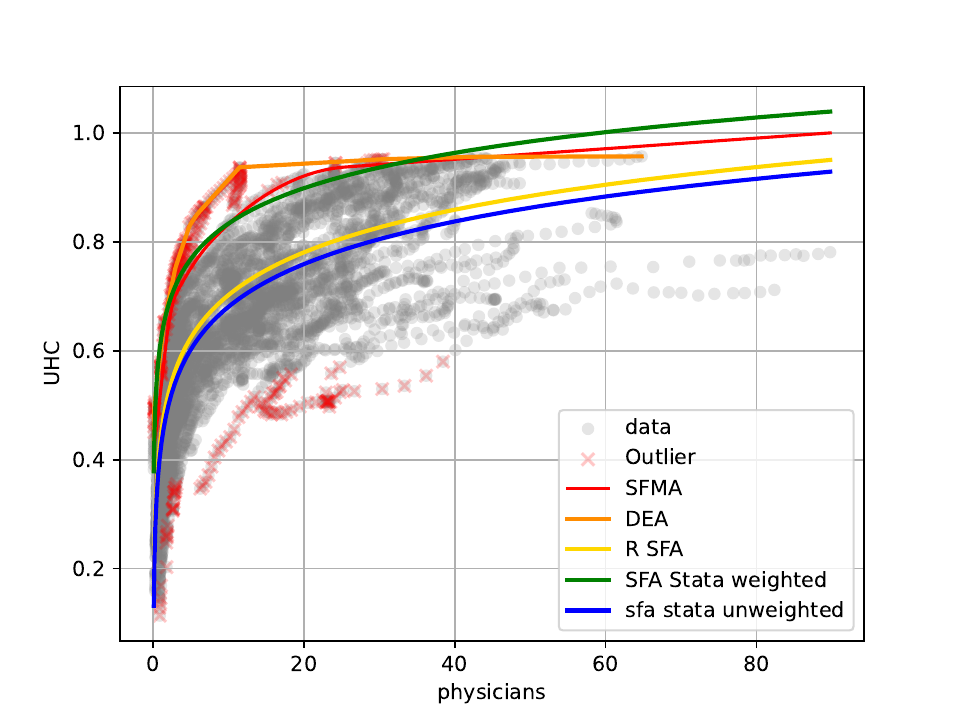}
\caption{Universal healthcare access (UHC) as a function of physicians per 10,000 population, log scale (top) and linear scale (bottom). DEA fits above all of the data. R SFA and default Stata SFA do  poorly on this datset, fitting through the middle of the data. Stata SFA with inverse variance weights gets a more useful frontier. SFMA is able to obtain the closest fit to the main cloud of data, also automatically trimming 5\% of observations. } \label{phys_uhc}
\end{center}
\end{figure}

\begin{figure}[h!]
\begin{center}
\includegraphics[scale=0.6]{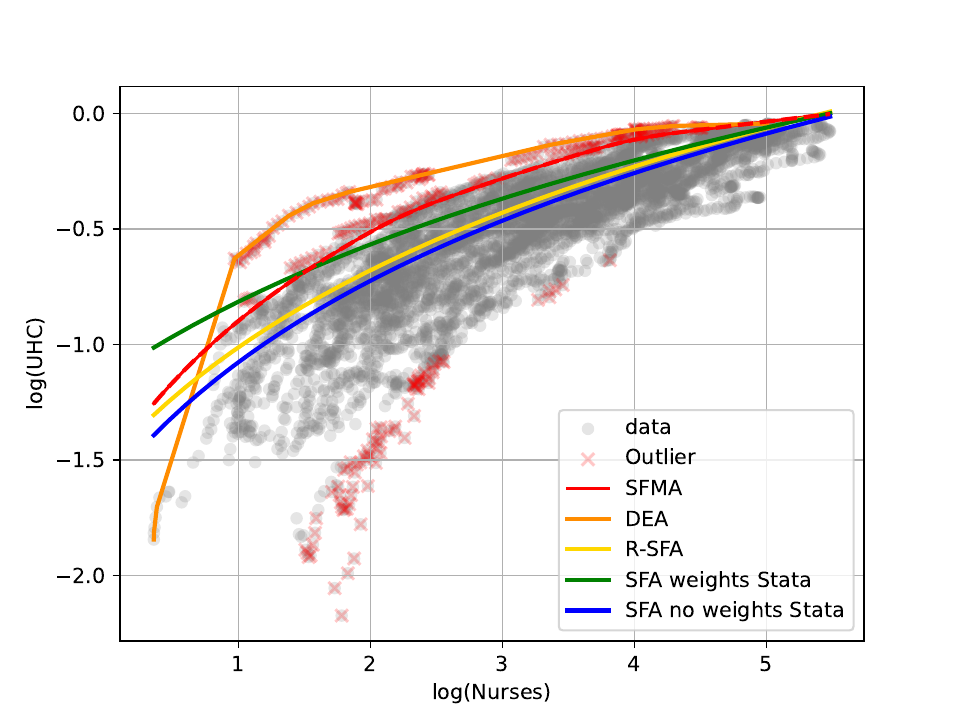}
\includegraphics[scale=0.6]{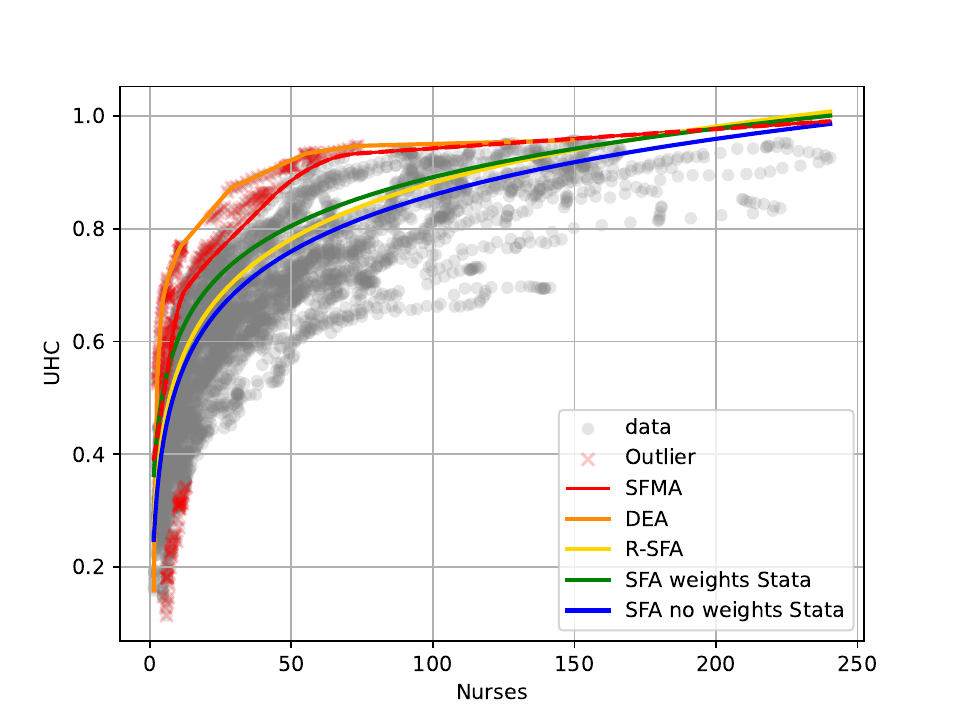}
\caption{Universal healthcare access (UHC) as a function of nurses per 10,000 population, log-scale (top) and linear scale (bottom).  DEA fits above all of the data. R SFA and both default and weighted Stata SFAs do poorly on this datset, fitting through the middle of the data. SFMA is able to obtain a closer fit to the cloud of data, also  automatically trimming 5\% of observations. } \label{nurse_uhc}
\end{center}
\end{figure}

 In fitting the frontiers, a few country-years with a low density of physicians score unusually well with respect to UHC. These cases confuse standard methods.  The SFA approach fits through the middle of the data. In an effort to obtain better frontier fits, we used the \verb{sfcross{ package in Stata. The package allows weighting, and for UHC examples, we indeed found that inverse variance weighting, using variances provided by the UHC dataset, produced a more useful result in the \verb{sfcross{ package than no weighting, as is seen in both linear and log scales in Figure~\ref{phys_uhc} and Figure~\ref{nurse_uhc}. Nonetheless, there is a tendency for all SFA fits to go through the middle of the data, just as the DEA result is guaranteed to go across the tops of the outlying datapoints, flattening the estimate of the underlying relationship. SFMA with trimming is able to get a useful fit to the majority of the data by automatically detecting and excluding 3\% of the data as outliers. We were not able to obtain fits for the StoNED package on this dataset, as the CNLS routine failed to converge.  Further details relating to the real data examples are presented in Section~\ref{sec:Discussion}.

\begin{table}[h!]
    \centering
    \caption{\label{tab:param_tab} List of Model Parameters}
    \begin{tabular}{l|c|c|c|c}\hline 
    Dataset  & No. knots  & Degree  & Trimming  & Constraints  \\ \hline
        GDP - LE            & 7      & 2      & 3\%     & Increasing, Concave  \\
        UHC-Physician           & 7      & 2       & 5\%      & Increasing, Concave  \\
        UHC-Nurse   & 7       & 2    & 5\%       & 
        Increasing, Concave
             \\\hline
    \end{tabular}
\end{table}

\section{Discussion}
\label{sec:Discussion}
We have developed a new semi-parametric approach to frontier estimation, dubbed SFMA.  The approach generalizes previous approaches to SFA, allowing for a semi-parametric frontier specification via splines with shape constraints, a flexible likelihood that accounts for both estimates and reported uncertainty, and can be used to automatically detect and remove outliers.  

To facilitate easy access to the new method we developed and implemented a customized optimization algorithm, which exploits the partially convex structure of the marginal likelihood function.

Numerical results show the utility of the modeling innovations. The value added of the innovations is clear as the simulations become more difficult, and trimming in particular is indispensable in the real applications, where ignoring small subsets of the data is essential to obtaining useful frontiers.

The applications to real datasets also demonstrate strengths of the modeling innovations. SMFA estimates a positive relationship between LE and GDP, with a steep increase from GDP per capita of \$148 to \$403 and more gradual from \$2981 to \$8103 than SFA. For example, in Vietnam, which is on the frontier from 2006 to 2011, LE increased from 74.3 in 2006 to 74.9 in 2011 when GDP per capita increased from \$784 to \$1,525 . In Spain, which is on the frontier in 2016 and 2019, LE increased LE from 83.32 to 83.49 when GDP per capita increased from \$26,523. to \$29,564. The functional form of SFA imposes a well-known, nearly vertical relationship between LE and GDP at lowest levels of GDP, but accurate data don't exist in this range of GDP, and the relationship is well below the frontier in the range where data exist. DEA estimates of the relationship as nearly flat, which is also contrary to observed increases.

SMFA estimates a relatively steeper increase in UHC from 0 to 10 physicians per 100,000 population, and more gradual increase above 20 physicians. Data on the frontier in the lowest range of physicians per 100,000 population represent countries such as Maldives with UHC score of 52.2 in 1990 and 0.9 physicians, 53.79 in 1992 and 1.05 physicians. Data on the frontier in the middle range of physicians per 100,000 population represent countries such as Iceland with a UHC score of 95.6 to 95.2 in 2011 and 2016, and 42.9 and 39 physicians, respectively. Three countries with 60 or more physicians per 100,000 population were well-below the frontier: Bermuda in 2009 and 2011, Cuba from 2004 to 2017, and Georgia from 2006 to 2017.

The results suggest that there are other factors associated with increases in UHC besides the number of physicians per capita. SFMA estimates a relatively steeper increase in UHC from 0 to 10 nurses per 100,000 population, and more gradual increase above 60 nurses.
Data on the frontier in the lowest range of nurses per 100,000 population represent countries such as Palestine with UHC score of 60 in 1991 and 1992 and 5.91 and 6.04 nurses respectively, and 62 in 1998 with 7.35 nurses. Data on the frontier in the middle range of nurses per 100,000 population represent countries such as Tunisia with a UHC score of 74 in 2005 and 17.47 nurses, 77 in 2013 and 23.90 nurses, 78 in 2014 and 2015, and 24.48, and 25.05 nurses, respectively. In the upper range, Luxemburg has a UHC score of 94 in 2016 and 2017, and 88.3 and 87.7 nurses per 100,000 population, respectively.

The methods developed here can be extended to any statistical models for inefficiencies that allows a closed form marginal likelihood, such as modeling inefficiencies using the Laplace distribution. A closed form marginal likelihood is leveraged to get the efficient algorithm that drives practical performance, leaving extensions where the marginal likelihood is not available in closed form to future work. 

\section{Acknowledgements} This work was funded by the Bill and Melinda Gates Foundation.



\bibliographystyle{abbrv}
\bibliography{meta}

\newpage
\section{Appendix}

\subsection{Derivative computations}

In this section we compute gradient and hessian of the half-normal SFMA likelihood. 
Recall the expression:
\begin{equation}
\label{eq:half_normal_model}
	\ell(\beta, \eta, \gamma) = \sum_{i=1}^m
		\frac{r_i^2}{2 v_i} +
		\frac{1}{2}\ln(v_i) -
		\ln\erfc(z_i r_i) = \sum_{i=1}^m
		\ell_i(\beta, \eta, \gamma)
\end{equation}
where,
\[
	r_i = y_i - \ip{x_i, \beta},\quad
	v_i = \sigma_i^2 + \eta + \gamma,\quad
	\tau_i = \sigma_i^2 + \gamma, \quad
	z_i = \sqrt{\frac{\eta}{2 \tau_i v_i}}.
\]
We first compute all  partial derivatives with respect to $\beta$, $\eta$ and $\gamma$.
\[
    \frac{\partial r_i}{\partial \beta} = -x_i, \quad
	\frac{\partial v_i}{\partial \eta} = 1, \quad
	\frac{\partial v_i}{\partial \gamma} = 1, \quad
	\frac{\partial \tau_i}{\partial \gamma} = 1.
\]
Next for $z_i$, which only depends on $\gamma$ and $\eta$, we have
\[
	\frac{\partial z_i}{\partial \eta} =
		\frac{z_i}{2}\lt(\frac{1}{\eta} - \frac{1}{v_i}\rt), \quad
	\frac{\partial z_i}{\partial \gamma} =
		\frac{z_i}{2}\lt(-\frac{1}{\tau_i} - \frac{1}{v_i}\rt).
\]
And for the second order derivatives we have,
\[\begin{aligned}
	\frac{\partial^2 z_i}{\partial \eta^2} &=
		\frac{z_i}{4}\lt(\frac{1}{\eta} - \frac{1}{v_i}\rt)^2 +
		\frac{z_i}{2}\lt(-\frac{1}{\eta^2} + \frac{1}{v_i^2}\rt),\\
	\frac{\partial^2 z_i}{\partial \eta \partial \gamma} &=
		\frac{z_i}{4}\lt(\frac{1}{\eta} - \frac{1}{v_i}\rt)
		\lt(-\frac{1}{\tau_i} - \frac{1}{v_i}\rt) +
		\frac{z_i}{2}\lt(\frac{1}{v_i^2}\rt),\\
	\frac{\partial^2 z_i}{\partial \gamma^2} &=
		\frac{z_i}{4}\lt(-\frac{1}{\tau_i} - \frac{1}{v_i}\rt)^2 +
		\frac{z_i}{2}\lt(\frac{1}{\tau_i^2} + \frac{1}{v_i^2}\rt).
\end{aligned}\]

The gradient and Hessian with respect to $\beta$ for a single term in the negative log likelihood sum is given by
\[\begin{aligned}
    \frac{\partial \ell_i(\beta, \eta, \gamma)}{\partial \beta} &=
		\lt(-\frac{r_i}{v_i} + h'(z_i r_i) z_i\rt) x_i, \\
	\frac{\partial^2 \ell_i(\beta, \eta, \gamma)}{\partial \beta^2} &=
		\lt(\frac{1}{v_i} - h''(z_i r_i) z_i^2\rt) x_i x_i^\top
\end{aligned}\]
%
The first derivatives with repsect to $\eta$ and $\gamma$ are given by
\[\begin{aligned}
	\frac{\partial \ell_i(\beta, \eta, \gamma)}{\partial \eta} &=
		-\frac{r_i^2}{2 v_i^2} + \frac{1}{2 v_i} -
		h'(z_i r_i) r_i \frac{\partial z_i}{\partial \eta},\\
	\frac{\partial \ell_i(\beta, \eta, \gamma)}{\partial \gamma} &=
		-\frac{r_i^2}{2 v_i^2} + \frac{1}{2 v_i} -
		h'(z_i r_i) r_i \frac{\partial z_i}{\partial \gamma}.
\end{aligned}\]
The second  derivatives are given by
\[\begin{aligned}
	\frac{\partial^2 \ell_i(\beta, \eta, \gamma)}{\partial \eta^2} &=
		\frac{r_i^2}{v_i^3} - \frac{1}{2 v_i^2} -
		h''(z_i r_i) r_i^2 \lt(\frac{\partial z_i}{\partial \eta}\rt)^2
		-h'(z_i r_i) r_i \frac{\partial^2 z_i}{\partial \eta^2},\\
	\frac{\partial^2 \ell_i(\beta, \eta, \gamma)}
		{\partial \eta \partial \gamma} &=
		\frac{r_i^2}{v_i^3} - \frac{1}{2 v_i^2} -
		h''(z_i r_i) r_i^2 \frac{\partial z_i}{\partial \eta}
		\frac{\partial z_i}{\partial \gamma}
		-h'(z_i r_i) r_i \frac{\partial^2 z_i}
		{\partial \eta \partial \gamma},\\
	\frac{\partial^2 \ell_i(\beta, \eta, \gamma)}{\partial \gamma^2} &=
		\frac{r_i^2}{v_i^3} - \frac{1}{2 v_i^2} -
		h''(z_i r_i) r_i^2 \lt(\frac{\partial z_i}{\partial \gamma}\rt)^2
		-h'(z_i r_i) r_i \frac{\partial^2 z_i}{\partial \gamma^2}.
\end{aligned}\]
\[\begin{aligned}
	\frac{\partial^2 \ell_i(\beta, \eta, \gamma)}
		{\partial \beta \partial \eta} &=
		\lt(\frac{r_i}{v_i^2} +
		h''(z_i r_i) z_i r_i \frac{\partial z_i}{\partial \eta} +
		h'(z_i r_i) \frac{\partial z_i}{\partial \eta}\rt) x_i, \\
	\frac{\partial^2 \ell_i(\beta, \eta, \gamma)}
		{\partial \beta \partial \gamma} &=
		\lt(\frac{r_i}{v_i^2} +
		h''(z_i r_i) z_i r_i \frac{\partial z_i}{\partial \gamma} +
		h'(z_i r_i) \frac{\partial z_i}{\partial \gamma}\rt) x_i.
\end{aligned}\]

\subsection{Basis Splines}
\label{sec:appendix-splines}

For general background on splines and spline regression see~\cite{de1978practical} and~\cite{friedman1991multivariate}.
For a spline with polynomial degree $p$ and $k$ knots placed at locations along the domain $x$, $t_0, ..., t_k$, we need $p + k$ basis elements to construct the spline basis, denoted $s^p_{j}$. These basis elements can be constructed recursively.

The $j$th element over domain $c_j^i$ is denoted by  $s_j^i$, with $1 \le j \le i + k$.

The spline basis is generated recursively:
By default we assume $c_0^i = \emptyset$, $c_{i+k+1}^i = \emptyset$ and $s_0^i(t) = 0$, $s_{i+k+1}^i(t) = 0$, for all $i \ge 0$.

\begin{figure}[ht]
\centering
\includegraphics[width=0.99\textwidth]{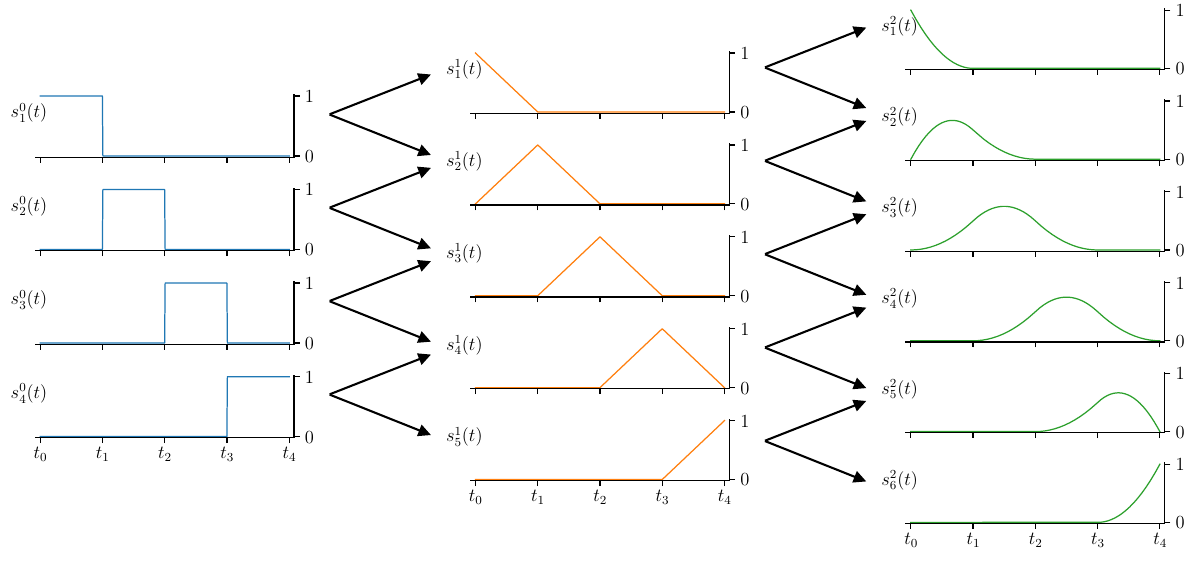}
\caption{Generation of bspline basis elements (orders 0, 1, 2).}
\label{fig:bspline}
\end{figure}

\begin{itemize}
\item $i = 0$,
\[
c_j^0 = [t_{j-1}, t_j),\quad
s_j^0(t) = \delta_{c_j^0}(t), \quad j = 1, \ldots, k
\]
\item $i > 0$,
\[
c_j^i = c_{j-1}^{i-1} \cup c_j^{i-1}, \quad s_j^i(t) = s_{j-1}^{i-1}(t) l(t; c_{j-1}^{i-1}) + s_j^{i-1}(t) r(t; c_j^{i-1}), \quad j = 1, \ldots, i+k
\]
where
\[
\begin{aligned}
l(t; c) &= \begin{cases}
(t - \inf c)/(\sup c - \inf c), & c \ne \emptyset\\
0, & c = \emptyset
\end{cases}, \\
r(t; c) &= \begin{cases}
(t - \sup c)/(\inf c - \sup c), & c \ne \emptyset\\
0, & c = \emptyset
\end{cases}.
\end{aligned}
\]
\end{itemize}
as illustrated in Figure~\ref{fig:bspline}.

With knowledge of the degree and knots placement, we can construct a \textit{design matrix} $X$ based on an input vector $x$, where the $j^{th}$ column of the design matrix is given by the expression
\begin{align}
\label{eq:splineDesign}
X_{\cdot, j} = 
    \begin{bmatrix} s_j^p(t_0) \\
                    \vdots \\
                    s_j^p(t_k)
    \end{bmatrix}.
\end{align}


\subsubsection{Linear Tails}
\label{sec:appendix-linear-tails}

We allow an option to make the first and/or last segment of the spline linear. The prior capabilities can then be used to 
set a prior for the slope of this segment to be $0$ (i.e. flat).  The estimated spline is then a best fit to the data, subject to this specification.
\begin{figure}[h]
\centering
\includegraphics[width=0.45\textwidth]{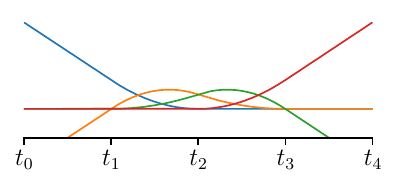}
\includegraphics[width=0.45\textwidth]{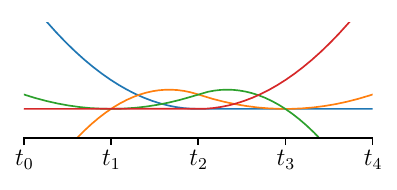}
\caption{Spline extrapolation. Left: linear extrapolation. Right: nonlinear extrapolation.}
\label{fig:bspline_extrapolation}
\end{figure}

Given knots $t_0, \ldots, t_k$, with $k \ge 3$, we want linear spline functions 
in domains $[t_0, t_1)$ and $[t_{k-1}, t_k)$. We implement this extension in two steps: 
\begin{itemize}
\item Generate elements $s_j^p(t)$ for the interval $[t_1, t_{k-1})$ with knots $t_1, \ldots, t_{k-1}$ and degree $p \ge 1$.
\item Extend the basis linearly to $[t_0, t_1)$ and $[t_{k-1}, t_k)$.
\end{itemize}

The key observation is that only $s_1, s_2$ have nonzero derivative at $t_1$, 
and only $s_{k+p-3}, s_{k+p-2}$ have nonzero derivative at  $t_{k-1}$:
\[
\begin{matrix}
{\displaystyle}
s_1^p(t_1) = 1 & \frac{d}{dt}s_1^p(t_1) = -\frac{p}{t_2 - t_1} & s_{k+p-2}(t_{k-1}) = 1 & \frac{d}{dt}s_{k+p-2}(t_{k-1}) = \frac{p}{t_{k-1} - t_{k-2}}\\
s_2^p(t_1) = 0 & \frac{d}{dt}s_2^p(t_1) = \frac{p}{t_2 - t_1} & s_{k+p-3}(t_{k-1}) = 0 & \frac{d}{dt}s_{k+p-3}(t_{k-1}) = -\frac{p}{t_{k-1} - t_{k-2}},
\end{matrix}
\]
Therefore we extend our spline to the first and last piece by 
\[
\begin{aligned}
\bar s_j^p(t) &= \begin{cases}
s_j^p(t_1) + \frac{d}{dt}s_j^p(t_1)(t - t_1), & t \in [t_0, t_1)\\
s_j^p(t), & t \in [t_1, t_{k-1})\\
s_j^p(t_{k-1}) + \frac{d}{dt}s_j^p(t_{k-1})(t - t_{k-1}), & t \in [t_{k-1}, t_k)
\end{cases}, \\
\frac{d}{dt} \bar s_j^p(t) &
= \begin{cases}
\frac{d}{dt}s_j^p(t_1), & t \in [t_0, t_1)\\
\frac{d}{dt}s_j^p(t), & t \in [t_1, t_{k-1})\\
\frac{d}{dt}s_j^p(t_{k-1}), & t \in [t_{k-1}, t_k)
\end{cases}.
\end{aligned}
\]
In general, using linear head and/or tail pieces to extrapolate outside the original domain or interpolate in the data sparse region is far more stable that using higher 
order polynomials, see Figure~\ref{fig:bspline_extrapolation}. 

\subsubsection{Shape Constraints}

We can use constraints to enforce monotonicity, convexity, and concavity. Monotonicity across the domain of interest follows 
from monotonicity of the spline coefficients. This relationship is derived for 
particular basis constructions by~\cite{de1978practical}, and has been used in the literature 
to enforce shape constraints~\cite{pya2015shape}. Current approaches work around the natural inequality constraints 
 by using  additional `exponentiated' variables.  Instead we impose these constraints directly as described below.  

Focusing just on $\alpha$, the relationship $\alpha_1 \leq \alpha_2$ can be written as $\alpha_1 - \alpha_2 \leq 0$. Stacking these inequality 
constraints for each pair $(\alpha_i, \alpha_{i+1})$ we can write all constraints simultaneously as 
\[
\underbrace{\begin{bmatrix}
1 & -1 & 0 & \dots & 0 \\
0 & 1 & -1 & \dots &  0 \\
\ddots & \ddots & \ddots &\ddots & \vdots\\
0 & \dots &\dots & 1 & -1
\end{bmatrix}}_{\bm C}
\begin{bmatrix} \alpha_1 \\ \alpha_2\\ \alpha_3 \\ \vdots \\ \alpha_n \end{bmatrix} 
\leq 
\begin{bmatrix} 0 \\ 0\\ \vdots \\ 0\end{bmatrix}. 
\]
These constraints are directly imposed through the optimization interface, along with any lower- and upper-limit constraints on $\alpha$. 

\paragraph{Convexity and Concavity}

For any $\mathcal{C}^2$ (twice continuously differentiable) function $f: \mathbb{R} \rightarrow \mathbb{R}$, convexity and concavity are captured by the 
signs of the second derivative. Specifically, $f$ is convex if $f''(t) \geq 0$ is everywhere, an concave if $f''(t) \leq 0$ everywhere. 
We impose linear inequality constraints on the expressions for $f''(t)$ over each interval. 
We can therefore easily pick any of the eight shape combinations given in~\cite[Table 1]{pya2015shape},
as well as imposing any other constraints on $\alpha$ (including bounds).

\subsection{Proof of Theorem~\ref{thm:h_concave}}

We first establish the global concavity of $\ln\erfc$.
For convienience, we denote $h := \ln \circ \erfc$. We have
\begin{equation}
\label{eq:erfcFormulas}
    \begin{aligned}
\erfc(x) &= 1 - \frac{2}{\sqrt{\pi}}\int_0^x \exp(-t^2)\,\mathrm{d}t\\
\erfc'(x) &= -\frac{2}{\sqrt{\pi}} \exp(-x^2)\\
\erfc''(x) &= \frac{4 x}{\sqrt{\pi}} \exp(-x^2)
\end{aligned}
\end{equation}
And we know,
\[
	h(x) = \ln\erfc(x), \quad
	h'(x) = \frac{\erfc'(x)}{\erfc(x)}, \quad
	h''(x) = \frac{\erfc''(x)\erfc(x) - \erfc'(x)^2}{\erfc(x)^2}
\]

We first consider the case $x \le 0$.
From the formulas~\eqref{eq:erfcFormulas}, we know that $\erfc$ is a positive and decreasing, and 
$\erfc''(x) \le 0$ when $x \le 0$.
Therefore,
\[
h''(x) = \frac{\erfc''(x) \erfc(x) - \erfc'(x)^2}{\erfc(x)^2} \le 0, \quad
\text{when } x \le 0.
\]

Next, we consider the case when $x > 0$. 
Here we use the asymptotic expansion of $h$ at $\infty$,
\[
\erfc(x) = \frac{2}{\sqrt{\pi}}\exp(-x^2)
\lt(
	\frac{1}{2x} + 
	\sum_{k=1}^\infty
	\frac{(-1)^k (2k - 1)!!}{2^{k + 1}x^{2k + 1}}
\rt)
\]
By grouping terms in the infinite sum in consecutive pairs, we see every pair is non-positive, which means the entire infinite sum is less than or equal to $0$. 
Therefore we have
\[
\erfc(x) \le \frac{2}{\sqrt{\pi}}\frac{\exp(-x^2)}{2x} \quad
\Rightarrow \quad
\erfc''(x) \erfc(x) - \erfc'(x)^2 \le 0
\]
From this, we immediately have that $h''(x) \le 0$, when $x > 0$. Combining with the previous case, we have $h''(x) \le 0$ for all $x \in \mathbb{R}$, and $h$ must be concave.

Next, the approximation for large values of $x$ is the sum of two terms: 
\[
\left(-x^2 - \ln(\sqrt{\pi}x)\right) + \ln\left(1-\frac{1}{2x^2}\right)
\]
The second derivative of the first term is given by $-2 +\frac{1}{x^2}$ and is negative for any $x > \frac{1}{\sqrt{2}}$. 
The second derivative of the second term is 
\[
\frac{-8x^2-4}{(2x^2-1)^2} + \frac{2}{x^2} \approx -\frac{2}{x^2}
\]
and is always negative. Thus the approximation is concave for 
 $x > \frac{1}{\sqrt{2}}$. 

\subsection{Interior Point Method for Smooth Convex Functions with Constraints}
\label{sec:interior} 

Consider the problem 
\[
\min_x f(x) \quad \mathrm{s.t.} \quad \ Cx \leq c. 
\]
where $f$ is twice continuously differentiable and convex. We introduce slack variables $s$ 
such that 
\[
C x + s = c. 
\]
The Lagrangian for the problem using the slack variables is given by 
\[
\mathcal{L}(x, s, \lambda) = 
f(x) + \lambda^\top(Cx + s - c),
\]
and at the optimal point, complementarity slackness conditions yield $\lambda_i s_i = 0$ for all $i$. 
The Karush-Kuhn-Tucker (KKT) optimality conditions are given by  
\[
0 = F\left(s, \lambda, x\right) = 
\begin{bmatrix}
Cx + s - c \\
\lambda \odot s \\
\nabla f(x) + C^\top\lambda
\end{bmatrix}
\]
The interior point update is obtained by forcing the constraints to be strictly satistfied, using a log-barrier method or, equivalently, modifying the nonlinear complementarity portion of the KKT condition to be strict with respect to a parameter $\mu$: 
\[
 F_\mu\left(s, \lambda, x\right) = 
\begin{bmatrix}
Cx + s - c \\
\lambda \odot s - \blue{\mu 1} \\
\nabla f(x) + C^\top\lambda
\end{bmatrix}
\]
We then apply a damped Newton iteration to drive $F_\mu$ to $0$, with each iteration given by 
\[
\underbrace{
\begin{bmatrix}
I & 0 & C \\
\Lambda & S & 0 \\
0 & C^\top & H(x) 
\end{bmatrix}
}_{\nabla F_{\mu}}
\begin{bmatrix}
\Delta s \\ \Delta \lambda \\ \Delta x 
\end{bmatrix}
= - \underbrace{
\begin{bmatrix}
Cx + s - c \\
\lambda \odot s - \mu 1 \\
\nabla f(x) + C^\top\lambda
\end{bmatrix}
}
_{F_{\mu}}
\]
where $H(x)$ is the hessian $\nabla^2 f(x)$. 
The Newton update is implementable whenever $H(x)$ is positive semidefinite (which is always true for a twice differentiable convex $f$), and in fact we can obtain explicit expressions of the Newton update for each coordinate:
\[
\begin{aligned}
\Delta x & = (H(x)+ C^\top \Lambda S^{-1} C)^{-1}(-f_3 + C^\top S^{-1}(f_2 - \Lambda f_1)) \\
\Delta s & = -f_1 - C\Delta x \\
\Delta \lambda &= -S^{-1}(f_2 + \Lambda \Delta s)
\end{aligned}
\]
For each iteration, we want to find the step size $\alpha$ that keeps the dual and slack variables strictly positive, while also ensuring descent in the optimality condition: 
\[
\begin{aligned}
& \lambda^+(\alpha) > 0  \\
& s^+(\alpha) > 0 \\
&\|F_\mu^+(\alpha)\|_\infty \leq (1- \gamma\alpha) \|F_\mu\|_\infty 
\end{aligned}
\]
Here we use $\gamma = 0.01$, which makes the last inequality easy to satisfy for relatively large steps. Once we find the step $\alpha$, we take 
\[
\begin{aligned}
s^+(\alpha) & = s + \alpha \Delta s\\
\lambda^+(\alpha) &= \lambda + \alpha \Delta \lambda \\
x^+(\alpha) &= x + \alpha \Delta x\\
F_\mu^+ & = F_\mu(s^+, \lambda^+, x^+)
\end{aligned} 
\]
To make the condition $F_\mu$ approach the optimality conditions $F$ of the original problem, we  drive the $\mu$ parameter to $0$ using a simple homotopy technique: 
\[
\mu^+ = 0.1\frac{\langle s^+, \lambda^+\rangle} {\#(s^+)}.
\]
In other words, at each iteration, $\mu$ is taken to be a fraction of how far complementarity conditions are away from $0$ on average. Once the (nonlinear) complementary slackness conditions are satisfied, $F =0$ will hold, since the affine equations in $F$ and $F_\mu$ are satisfied in every Newton update.  

\subsection{StoNED formulation}

Consider the model
\[
\begin{aligned}
y_i &= f(x_i) - u_i - v_i 
\end{aligned}
\]
where $u_i > 0$ is the inefficiency term and $v_i$  is the noise term. We assume $u_i$ has mean $\mu > 0$ and variance $\sigma_u^2$, $v_i$  has mean $0$ and variance $\sigma_v ^2$.
The first stage in StoNED estimates the functional form of the frontier using CNLS:
\[
\begin{aligned}
\min_{v,\alpha,\beta} \sum_{i=1}^n v_i^2 \\
y_i=\alpha_i + \beta_i x_i + v_i \\
\alpha_i + \beta_i x_i \le \alpha_h + \beta _h x_i, \forall h, i=1,...,n \\
\beta_i \ge 0, \forall i=1,..,n \\
\end{aligned}
\] 
Once $v_i^2$ is obtained, there are different choices of mthods to obtain noise and inefficiency terms~\cite{kuosmanen2012stochastic}, as summarized below. 
\subsubsection{Method of Moments (MoM)}
 MoM assumes normality of sampling error and a half normal inefficiency term, and uses skewness formulas to properties to disentangle the noise from the inefficiency term \cite{fan1996semiparametric}.
Given the estimated second and third moment  $\hat{M_2}$ and $\hat{M_3}$of the  CNLS residuals, the variance parameters are given by~\cite{kuosmanen2012stochastic}.
\[
\begin{aligned}
\hat{\sigma_u} = \sqrt[3]{\frac{\hat{M_3}}{\left[\frac{2}{\pi}\right]\left[1 - \frac{4}{\pi}\right]}} \\
\hat{\sigma_v} = \sqrt{\hat{M_2} - [\frac{\pi - 2}{\pi}}]\hat{\sigma_u} \\
\end{aligned} \\ 
\]

\subsubsection{Pseudolikelihood Estimation (PSL):}
Estimating the standard deviations $\sigma_u$ and $\sigma_v$ can also be obtained using PSL, which was first coined by \cite{fan1996semiparametric}. Here, parameters are set as $\sigma = \sigma_u + \sigma_v$ , and obtained by maximizing log-likelihood as a function of $\lambda$ , which is given as 

\[
\begin{aligned}
\log L(\lambda)=-n\log\hat{\sigma} + \sum _{i=1}^n \log \phi [\frac{-\epsilon_i \lambda}{\hat{\sigma}}] - \frac{1}{2 \hat{\sigma}^2}\sum _{i=1}^n \hat{\epsilon_i}^2\\
\hat{\varepsilon }_{i} = \hat{\upsilon}_{i} - \frac{{\sqrt{2}\lambda\hat{\sigma}}}{\pi(1 + \lambda^2)^{1/2}} \\
\hat{\sigma} = [\frac{\frac{1}{n} \sum_{j=1}^n \hat\upsilon^2}{[1 - \frac{2\lambda^2}{\pi(1+\lambda)}]}]^{1/2} \\
\end{aligned}
\] 
Here $\phi$ is the standard normal cumulative distribution function.

\subsubsection{Estimation of inefficiency term}
Once an estimator of $\sigma_u$ is available from MoM or PSL,  the production frontier is obtained by shifting the CNLS estimate by a factor of $\hat{\sigma_u}\sqrt{\frac{2}{\pi}}$. \\
The conditional expected value of the inefficiency is given by 

\[
\begin{aligned}
\hat{E}(u_{i} \mid \hat{\varepsilon}_{i}) = -\frac{\hat{\varepsilon}_{i} \hat{\sigma}_{u}^{2}}{\hat{\sigma}_{u}^{2} + \hat{\sigma}_{v}^{2}} + \frac{\hat{\sigma}_{u}^{2} \hat{\sigma}_{v}^{2}}{\hat{\sigma}_{u}^{2} + \hat{\sigma}_{v}^{2}}\left[\frac{\phi(\hat{\varepsilon}_{i}/\hat{\sigma}_{v}^{2})}{1 - \Phi(\hat{\varepsilon}_{i}/\hat{\sigma}_{v}^{2})}\right]
\end{aligned}
\] \\
where $\hat{\varepsilon}_{i} = \hat{\upsilon}_{i} - \hat{\sigma}_{u} \sqrt{\frac{2}{\pi}}
$ is the estimator for the composite error term \cite{kuosmanen2012stochastic}.

\end{document}